\documentclass[aps,prd,twocolumn,showpacs,superscriptaddress,nofootinbib,a4paper,10pt, reprint,eqsecnum,reprint]{revtex4-2}

\usepackage{newtxmath,newtxtext}
\usepackage{mathtools}
\usepackage{bm}
\usepackage{graphicx}
\usepackage{dcolumn}
\usepackage[dvipsnames]{xcolor}
\usepackage{hyperref}
\usepackage{booktabs}
\usepackage{multirow}
\usepackage{makecell}
\usepackage{orcidlink}

\hypersetup{colorlinks=true,linkcolor=blue,citecolor=ForestGreen,urlcolor=blue}

\newcommand{\cs}{\ensuremath{c_{\rm s}^{2}}}
\newcommand{\ca}{\ensuremath{c_{\rm a}^{2}}}

\newcommand{\nb}{\ensuremath{ \nabla }}
\newcommand{\m}{\ensuremath{{\mu \nu}}}

\newcommand{\rde}{\ensuremath{\rho_{\rm de}}}

\newcommand{\km}{\ensuremath{\rm km \ s^{-1} Mpc^{-1}}\ }

\begin{document}
	
	\title{Observational constraints on a damped harmonic oscillator model of dark energy}
	
	\author{Saddam Hussain\orcidlink{0000-0001-6173-6140} }
	\email{saddamh@zjut.edu.cn (Corresponding Author)}

	\affiliation{Institute for Theoretical Physics and Cosmology, Zhejiang University of Technology,
		Hangzhou 310023, China}
	
	\author{Muhammad Ahmad}
	\email{ahmad@xao.ac.cn (Corresponding Author)}
	\affiliation{State Key Laboratory of Radio Astronomy and Technology, Xinjiang Astronomical Observatory, CAS, 150 Science 1-Street, Urumqi 830011, China }
	
	\author{Ming Zhang}
	\email{zhangm@xao.ac.cn}
	\affiliation{State Key Laboratory of Radio Astronomy and Technology, Xinjiang Astronomical Observatory, CAS, 150 Science 1-Street, Urumqi 830011, China }
	
	\author{Ayodeji Ibitoye}
	\email{ayodeji.ibitoye@gtiit.ac.cn}
	\affiliation{Department of Physics, Guangdong Technion – Israel Institute of Technology, Shantou, Guangdong 515063, People’s Republic of China}
	\affiliation{Centre for Space Research, North-West University, Potchefstroom 2520, South Africa }
	\affiliation{Department of Physics and Electronics, Adekunle Ajasin University, P. M. B. 001, Akungba-Akoko, Ondo State, Nigeria}
	
	\author{Lang Cui}
	\email{cuilang@xao.ac.cn (Corresponding Author)}
	\affiliation{State Key Laboratory of Radio Astronomy and Technology, Xinjiang Astronomical Observatory, CAS, 150 Science 1-Street, Urumqi 830011, China }
	
	\date{\today}
	
	\begin{abstract}
		We constrain a damped harmonic oscillator (DHO) dark-energy equation of state using the full cosmic microwave background (CMB) likelihoods in combination with DESI BAO and three distinct Type~Ia supernova compilations: Pantheon+, DES-Dovekie, and Union3. The equation of state obeys a second-order damped oscillator equation in number of $e$-folds, so that its frequency $f$, damping rate $b$ and equilibrium value $w_{\rm m}$ fully specify the late-time dynamics. The model exhibits oscillatory behavior only at low redshifts, around the equilibrium value $w=-1$, with distinct characteristics for the different supernova compilations: an underdamped solution for DES-Dovekie and Union3, and an overdamped solution for Pantheon+. At higher redshifts, the model closely mimics $\Lambda$CDM and deviates significantly only at $z<0.6$, with the magnitude of the deviation depending on the supernova compilation. We further identify a region of the $(f,b)$ parameter space, corresponding to rapid variation of the equation of state at low redshift, in which the
		perturbation equations become numerically stiff and cannot be integrated with a
		canonical dark-energy sound speed i.e., $\cs =1$. We show that reducing the rest-frame sound
		speed removes this obstruction while leaving the observables unchanged at the
		$10^{-3}$ level, and therefore treat it as a numerical prescription rather than a physical modification of the model. The model yields $H_0 = 67.53^{+1.22}_{-1.18}$~km/s/Mpc for Pantheon+, $H_0 = 69.08^{+1.23}_{-1.16}$~km/s/Mpc for DES-Dovekie, and $H_0 = 70.67^{+1.89}_{-1.87}$~km/s/Mpc for Union3. The present-day equation-of-state parameter is constrained to $w_0 = -0.521^{+0.891}_{-0.414}$, $-3.01^{+1.14}_{-1.21}$, and $-3.16^{+1.11}_{-1.16}$ for Pantheon+, DES-Dovekie, and Union3, respectively. The difference in the minimum chi-squared, $\Delta\chi_{\rm min}^2$, marginally favors the CPL model over the damped harmonic oscillator (DHO) model for the Union3 and DES-Dovekie compilations, while the logarithmic Bayes factor consistently favors CPL over DHO, with the strength of the evidence depending on the supernova compilation.
	\end{abstract}
		
	\maketitle
	
	\section{Introduction}
	
	Our understanding of cosmology has been profoundly challenged and continues to
	evolve with the unprecedented precision of {modern} cosmological
	observations. Observations including Type Ia Supernovae
	\cite{SupernovaCosmologyProject:1998vns,SupernovaSearchTeam:1998fmf}, the Cosmic
	Microwave Background {(CMB)}
	\cite{Baccigalupi:2006ww,Sherwin:2011gv,Planck:2015fie,Planck:2018vyg},
	Baryon Acoustic Oscillations {(BAO)}
	\cite{Eisenstein:2005sbt,Percival:2007yw,BOSS:2016wmc}, and weak lensing surveys
	\cite{Li:2008vf,DES:2017tss,LSST:2008ijt} have
	{provided compelling evidence for the existence of} two dark
	components, known as dark energy and dark matter. Despite their dominant
	contributions to the cosmic energy budget, the physical nature and origin of
	these components remain {unknown} and constitute well-known open problems in {cosmology}. The simplest model that has been widely
	accepted is {the} $\Lambda$CDM {model}, as it
	provides an excellent fit to a broad range of cosmological observations, but it
	is also criticized due to persistent theoretical inconsistencies, including the
	cosmological coincidence problem and fine-tuning issues
	\cite{Weinberg:1988cp,Zlatev:1998tr}. {Here $\Lambda$ denotes} 
	the cosmological constant, which {drives accelerated expansion} through its negative pressure, $P = -\rho_{\Lambda}$, with a positive energy
	density. It contributes roughly $69\%$ of the total energy budget of the
	Universe, while the other major component, contributing $\sim 25\%$, is cold dark
	matter (CDM), which {does not interact electromagnetically} and behaves as pressureless, non-relativistic matter
	\cite{Copeland:2006wr,Sahni:1999gb,Sahni:2004ai,Frieman:2008sn}. The remaining energy budget consists of non-relativistic baryonic matter, which accounts for
	about ${5}\%$, while radiation contributes
	{only $\lesssim 1\%$ of the present-day total}.
	
	Further improvements in the precision of cosmological observations have revealed
	several notable issues, including the discrepancy in the measured value of the
	Hubble constant $H_0$ at the $\sim 5\sigma$ level between
	{Planck} \cite{Planck:2018vyg} and SH0ES
	\cite{Riess:2021jrx,Breuval:2024lsv}. Furthermore, the release of
	{BAO} measurements from the Dark Energy Spectroscopic Instrument
	(DESI) \cite{DESI:2024mwx} has shown a preference for dynamical dark energy,
	described by the $w_0w_a${CDM} (CPL) model
	{\cite{Chevallier:2000qy,Linder:2002et}}, over the cosmological
	constant at the $2.6\sigma$ level for DESI+CMB, with the significance increasing
	to $3.9\sigma$ depending on the inclusion of the Pantheon+ \cite{Brout:2022vxf},
	Union3 \cite{Rubin:2023jdq}, or DES-SN5YR \cite{DES:2024upw} supernova
	compilations. The subsequent release of updated DESI BAO measurements
	\cite{DESI:2025zgx} increased the preference for dynamical dark energy to
	$3.1\sigma$ for DESI+CMB, while the significance varies between $2.8$ and
	$4.2\sigma$ depending on the supernova compilation used. Independent groups have
	investigated this apparent preference and suggested that it could be associated
	with systematic effects in the low- and high-redshift samples
	\cite{Efstathiou:2024xcq,Ormondroyd:2025phk}. These concerns have been addressed
	in \cite{DES:2025tir}, which confirmed the results obtained with DES-SN5YR.
	However, a {recalibration} of the five-year Type Ia supernova
	sample, DES-SN5YR, {referred to as} DES-Dovekie
	\cite{DES:2025sig}, has {tightened} the constraints on the
	$w_0w_a$ parameters and reduced the preference for dynamical dark energy over the
	cosmological constant. {The} debate over whether the preference 	for dynamical dark energy originates from systematic effects or from progenitor-age bias associated with the supernova sample
	\cite{Sah:2026mgk,Chung:2025cgv,Wiseman:2026nty,chung_reply2025} {thus} remains far from settled. Future supernova observations from the Vera C. Rubin Observatory \cite{Simongini:2025hel} and the Nancy Grace
	Roman Space Telescope \cite{Kessler:2025eib} should help resolve this issue with greater precision.
	
	The evidence for dynamical dark energy is, however, model dependent, as there is
	currently no {observation that probes its equation of state
		directly}, nor do we have a fundamental theory from which {that equation of state} can be derived. {The
		CPL model should therefore not be treated as a benchmark} for probing the dynamical nature of dark energy (DE) merely on the grounds that it is one of the
	simplest extensions of the cosmological constant based on a Taylor series expansion. One {limitation} of this
	{parametrization} is that it cannot capture
	{non-monotonic behavior such as oscillations} \cite{Kessler:2025kju}. {In the absence of a unique equation of state,} a large number of parameterized phenomenological models have been proposed
	\cite{Efstathiou:1999tm,Mehrabi:2018oke,Yang:2017alx,Yang:2021flj,Pan:2019brc,Li:2019yem,Escamilla:2023oce,Escamilla:2024fzq,Hussain:2025nqy,Choudhury:2003tj,Gong:2005de,Jassal:2004ej}. Various physically motivated scalar-field models have also been proposed to
	explain the dynamical nature of dark energy and potentially address the cosmic coincidence problem
	\cite{Ratra:1987rm,Tsujikawa:2013fta,Armendariz-Picon:2000nqq,Armendariz-Picon:2000ulo,Chiba:1999ka,Kamenshchik:2001cp,Bento:2002ps,Hussain:2024qrd}. Alternative approaches include modifying the gravity sector
	\cite{Sotiriou:2008rp,Nojiri:2006ri,BeltranJimenez:2017tkd,Cai:2015emx,Avilez:2013dxa,SolaPeracaula:2019zsl,SolaPeracaula:2020vpg,Joudaki:2020shz,Myrzakulov:2025jpk,Hussain:2023kwk} and introducing interactions between the dark energy and dark matter sectors
	\cite{Ellis:1989as,Amendola:1999qq,Farrar:2003uw,Zimdahl:2003wg,Sadjadi:2006qp,Hussain:2022dhp,Das:2023rat,Amendola:1999er,Nunes:2000ka,Chimento:2003iea,Chatterjee:2021ijw,Hussain:2025vbo,Hussain:2025uye,Arora:2025ecj,Hussain:2024jdt}.
	
	Recent efforts based on numerical and statistical tools, which offer greater
	flexibility than conventional parameterized models, have revealed oscillatory
	features in the {dark energy equation of state}
	\cite{Zhao:2017cud,Ormondroyd:2025exu,DESI:2025wyn,Ormondroyd:2025iaf,Escamilla:2021uoj}
	at lower redshifts, motivating the
	construction of phenomenological models with an oscillatory equation of state
	\cite{Brown:2017osf,Li:2011dr,Pan:2017zoh,Rezaei:2024vtg}. Earlier studies also investigated oscillating dark energy as a possible 
	mechanism for addressing some of the important shortcomings of $\Lambda$CDM, including the cosmic coincidence problem
	\cite{Linder:2005dw,Rubano:2003er,Lazkoz:2007mx,Zhao:2006mn,Leon:2012mt,Das:2013sca,Pace:2011kb}.
	
	Given our limited understanding of the physical origin of dark energy, there is
	currently no compelling reason to exclude an oscillatory equation of state,
	provided that it remains consistent with observations. In earlier studies
	\cite{Rezaei:2024vtg,Kessler:2025kju}, {oscillatory dark energy was modelled using} ad hoc trigonometric functions, which do not arise from a
	common dynamical principle governing the oscillatory evolution. In such cases, it
	is difficult to characterize the properties of a generic oscillatory system, such
	as the roles of the frequency and damping factor. In light of this, Hussain {\textit{et al.}} \cite{Hussain:2026srf} proposed a common dynamical framework in which the dark energy equation of state satisfies a
	second-order damped harmonic {oscillator} equation. In this
	framework, the model oscillates around the equilibrium value $w_m$, with
	frequency $f$ and damping factor $b$, which uniquely determine the dynamical
	characteristics of the model and allow for underdamped, overdamped, or critically
	damped solutions. {Several} variants of the model have been
	studied using several combinations of observational data, including Planck
	distance priors. The variant with $w_m=-1$ prefers an underdamped solution for
	DES-Dovekie and Union3, whereas the Pantheon+ compilation yields a critically
	damped solution. Owing to its distinct dynamical
	{behavior across} supernova compilations, the model provides
	{a means of} probing the geometric information encoded in
	different supernova datasets. Moreover, this
	{model} yields higher Bayesian evidence for the Pantheon+
	compilation compared to both $\Lambda$CDM and CPL when combining late- and
	early-time datasets. Other variants {have been constructed with} different equilibrium points, $w_m=-0.8$ and $-1.2$, corresponding to the
	quintessential and phantom regimes{; both} lie in the overdamped regime{, and therefore exhibit} no oscillatory solution.
	Furthermore, the role of the damping factor has been investigated by setting
	$b=0$, for which the equation of state oscillates around $w_m=-1$ throughout the
	cosmic epoch with a constant frequency. However, in the absence of damping, such
	oscillations do not produce significant {features} in the
	evolution of the Hubble parameter or the deceleration parameter $q$, and the
	resulting effective background evolution remains close to that of $\Lambda$CDM.
	
The appeal of the damped harmonic oscillator (DHO) prescription lies in the fact that it is not merely a fitting function but a dynamical law. Once the equilibrium value $w_{\rm m}$ is fixed, the entire late-time evolution of the equation of state is determined by two physically interpretable quantities---the oscillation frequency $f$ and the damping rate $b$---together with two initial conditions. The qualitative character of the solution is governed by the discriminant, $b^{2}-4f^{2}$, which partitions the parameter space into underdamped, critically damped, and overdamped branches. This provides a considerably stronger framework than ad hoc trigonometric forms, as it associates each supernova compilation not merely with a set of best-fit parameters but with a distinct dynamical regime. It also embeds the cosmological constant as an asymptotic attractor toward the past rather than as a fine-tuned special case: for $b>0$, every solution approaches $w_{\rm de}=w_{\rm m}=-1$ as $N\rightarrow-\infty$, so the model naturally approaches $\Lambda$CDM at high redshift without any additional assumption. Any departure from $\Lambda$CDM is therefore confined to low redshift by construction, which is precisely where the current tension between the supernova compilations is most pronounced.
	
The analysis of Ref.~\cite{Hussain:2026srf} was carried out at the level of the background expansion, using compressed CMB distance priors. This is adequate for testing the geometric predictions of the model but does not address two important questions. First, distance priors compress the CMB information and are calibrated assuming a $\Lambda$CDM-like recombination history; consequently, they cannot capture the full impact of a dynamical dark-energy component on the temperature and polarization power spectra, particularly its effect on the late-time integrated Sachs--Wolfe (ISW) effect. Second, and more importantly, a background-only treatment does not address dark-energy perturbations, which, for an equation of state that repeatedly crosses the phantom divide, are neither trivial nor guaranteed to remain numerically well behaved. The present work addresses both issues. We implement the DHO equation of state directly in the Boltzmann solver \texttt{CLASS}, integrating the governing oscillator equation numerically alongside the background and perturbation equations, and confront the model with the full Planck likelihood.
	
	In doing so, we encounter a feature that we believe is of practical relevance to studies of rapidly varying dark-energy models: over an identifiable region of the $(f,b)$ plane, the linear perturbation system cannot be integrated reliably. We show that this numerically inaccessible region is controlled by the characteristic timescales of the oscillator modes and corresponds to cases in which the equation of state varies on a conformal timescale that becomes short compared with the characteristic timescale of the subhorizon dark-energy perturbations. We further show that the numerical obstruction can be alleviated by lowering the rest-frame sound speed, $\cs\to0$, and---crucially---that for the present class of models this modification has a negligible impact on the relevant cosmological observables. Thus, lowering $\cs$ serves primarily as a numerical prescription for accessing otherwise difficult regions of parameter space rather than as a modification that materially changes the physical predictions of the model.
	
	In this paper, we further study this model,
	{specifically the} variant {that oscillates}
	around the mean value $-1$ with {a} non-zero damping factor, and
	confront it with the full {CMB likelihood}
	\cite{Planck:2018vyg} in combination with DESI BAO \cite{DESI:2025zgx} and three
	{compilations} of Type Ia supernovae, namely
	Pantheon+ \cite{Brout:2022vxf}, DES-Dovekie \cite{DES:2025sig}, and Union3
	\cite{Rubin:2023jdq}.
	
	
	In Sec.~\ref{sec:background}, we introduce the dark energy equation of state
	parametrization and outline the basic dynamical equations in a homogeneous and
	isotropic background, together with the linear perturbation equations governing the dark-energy fluid. The observations and methodology are discussed
	in Sec.~\ref{sec:obs_method}, while a detailed discussion of the model's pathologies and constraints is presented in Sec.~\ref{sec:result}. Finally, we
	summarize our conclusions in Sec.~\ref{sec:conclusion}.

\section{Background dynamics}
\label{sec:background}

At the largest scales, assuming that the Universe is homogeneous and isotropic, the line element is described by the Friedmann--Lema\^{\i}tre--Robertson--Walker (FLRW) metric
\begin{equation}
	ds^2=-dt^2+a^2(t)\left[\frac{dr^2}{1-k r^2}+r^2\left(d\theta^2+\sin^2\theta\  d\phi^2\right)\right],
\end{equation}
where $a$ is the scale factor, and throughout this study we assume a spatially flat geometry, i.e., $k=0$, with $(t,r,\theta,\phi)$ denoting the comoving coordinates. Assuming that the matter content of the Universe is minimally coupled to gravity, i.e., $\nb_{\mu}T^{\m}=0$, where $T^{\m}$ is the stress-energy tensor that includes radiation (r), baryons (b), dark matter (dm), and dark energy (de), the continuity equation for the considered metric becomes
\begin{equation}
	\dot\rho_{i} + 3 H (\rho_i +P_i) = 0,
\end{equation} 
where an overdot denotes a derivative with respect to cosmic time, $\rho_{i}$ represents the energy density of the $\rm i^{th}$ species, and $P_{i}$ denotes its corresponding pressure. The ratio of the pressure to the energy density defines the equation of state (EOS) of the corresponding species as $w_{i} \equiv P_i/\rho_i$. The EOS of radiation is $1/3$, while baryons and dark matter behave as pressureless dust fluids. For dark energy, $w_{\rm de}=-1$ corresponds to a cosmological constant. In the present study, however, we assume that the dark energy EOS is time dependent and characterize it by the following ansatz:
\begin{eqnarray}
	\text{\rm \bf DHO:} & \frac{d^2 w_{\rm de}}{dN^2}
	= -f^2\bigl(w_{\rm de}-w_{\rm m}\bigr)
	+ b\,\frac{dw_{\rm de}}{dN},
	\label{eq:osc}
\end{eqnarray} 
where $z$ is the redshift and $N = \ln(a) = -\ln(1+z)$. The DHO model represents a second-order linear differential equation for the EOS as a function of the number of e-folds and is introduced to capture the oscillatory behavior of dark energy (DE)\footnote{The DHO model represents a particular subclass of the general class of second-order evolution equations for $w_{\rm de}(N)$ introduced in Ref.~\cite{Paliathanasis:2025cuc}.}. Here, $f$ denotes the frequency of the oscillation, $b$ characterizes the damping effect, and $w_m$ represents the equilibrium point of the oscillation, which we set to $w_m = -1$. To solve the equation, two initial conditions are required, namely $w_{\rm de}(0) = w_0$ and $w_a \equiv \frac{dw_{\rm de}}{dN}|_{N=0}$. Therefore, the model has a total of four degrees of freedom, which are sampled in the MCMC analysis. Hussain \textit{et al.} \cite{Hussain:2026srf} found that the posterior distribution of $w_a$ remains largely unconstrained for all combinations of datasets and, hence, does not significantly influence the posterior distributions of the other parameters. Nevertheless, for the sake of completeness, we vary all the model parameters in our analysis. 

Since Eq.~\eqref{eq:osc} is linear with constant coefficients and does not
couple directly to the expansion rate, it admits a closed-form solution. The
corresponding characteristic roots are
\begin{equation}
	\lambda_{\pm} = \tfrac{1}{2}\left(b \pm \sqrt{b^{2}-4f^{2}}\right),
	\label{eq:roots}
\end{equation}
and the general solution can be written as
\begin{equation}
	w_{\rm de} = w_{\rm m} + c_1 e^{\lambda_+ N} + c_2 e^{\lambda_- N}\ .
\end{equation}
For $b>0$, both modes decay toward the past, $N\to-\infty$, and hence the
equation of state approaches $w_{\rm m}$, which in the present case is $-1$.
An oscillatory solution is realized for $b<2f$, with angular frequency
$\omega=\sqrt{f^{2}-b^{2}/4}$, whereas the solution is purely exponential for
$b>2f$. Thus, for $b>0$, significant deviations from $\Lambda$CDM are naturally
confined to the late Universe. Toward the future, $N\to+\infty$, the amplitude
of the oscillatory solution for $0<b<2f$ grows exponentially, and the equation
of state can develop increasingly large deviations from $w_{\rm m}$. This
behavior can lead to a future singularity in the extrapolated cosmological
evolution. Such future pathologies are not unique to the DHO model; other
phenomenological parameterizations, including CPL, can also exhibit
unphysical behavior when extrapolated beyond the redshift range probed by
observations. Since the available observational data primarily constrain the
past light cone, the model remains well behaved over the observationally
relevant redshift range.

One may alternatively consider negative $b<0$, for which the solutions decay
toward the future, $N\to+\infty$, and $w_{\rm de}\to w_{\rm m}$. However, in
this case the oscillatory amplitude grows toward the past for the underdamped
branch and can therefore become excessively large at high redshift. We
therefore restrict the negative-$b$ range in Tab.~\ref{tab:prior} such that
the resulting oscillations do not produce a large dark-energy density that
could significantly affect the early Universe. Thus, the model can in
principle produce oscillatory behavior either toward the late or early
Universe, depending on the sign and magnitude of $b$, and the observationally
preferred regime can therefore depend on the cosmological probe.

\begin{figure*}[tbh]
		\resizebox{\linewidth}{!}{\includegraphics{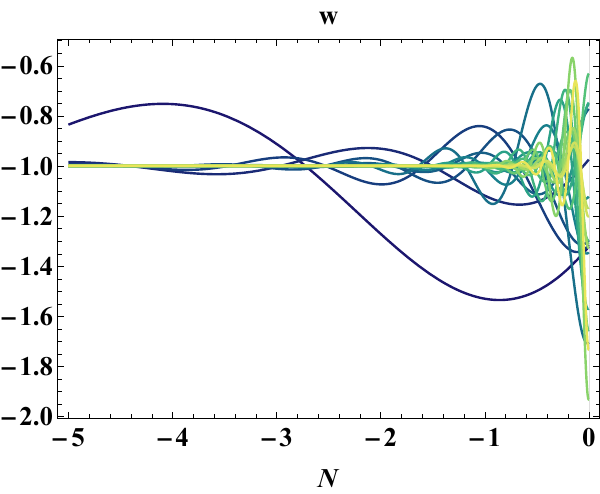}, \   \includegraphics{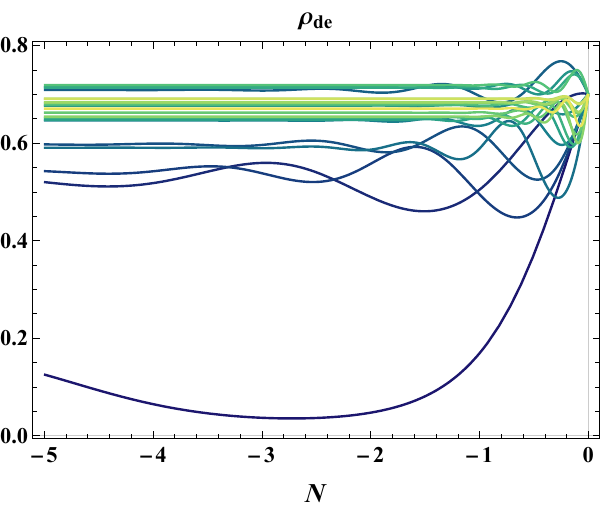}, \includegraphics{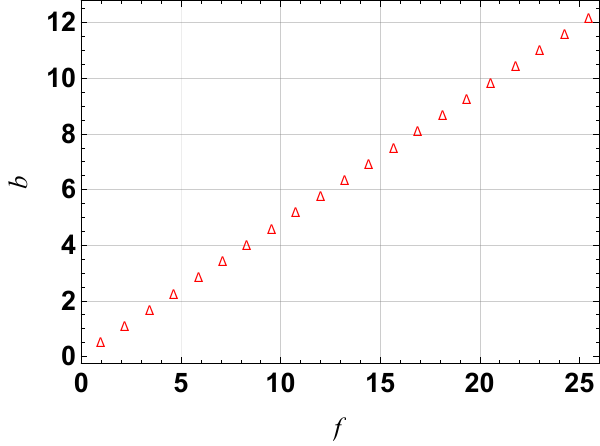}}
		\caption{Variation of the equation of state $w_{\rm de}$ and energy density $\rho_{\rm de}$ corresponding to DHO for $-2< w_0<-0.6$ and fixed $w_a = 0.5$, for distinct values of $f$ and $b$ shown on the right. The color coding is such that the lowest values of $f$ and $b$ are represented in blue; as their values increase, the curves transition to green, while higher combinations are represented by yellow curves. It can be seen that for higher frequencies, the energy density exhibits oscillations with lower amplitudes and eventually approaches a constant value, whereas for lower frequencies, the density fluctuations have larger amplitudes. Consequently, low-frequency oscillations can lead to a higher dark energy density, which may significantly alter the evolution of the early Universe.}
		\label{fig:osc_eos_numeric_variation}
	\end{figure*}

Nevertheless, Ref.~\cite{Hussain:2026srf} showed that the potential future
pathology with $b>0$ can be avoided by replacing $b$ with $b\tanh(-\gamma N)$, such that
the dynamical equation of state becomes
\begin{equation}
	\frac{d^{2} w_{\rm de}}{dN^{2}}
	= -f^{2}\bigl(w_{\rm de}-w_{m}\bigr)
	+ b \tanh(-\gamma N)\,\frac{dw_{\rm de}}{dN}\,.
	\label{eq:osc2}
\end{equation}
However, in the present work, we restrict our analysis to the original DHO prescription specified in Eq.~\eqref{eq:osc}.
Fig. \ref{fig:osc_eos_numeric_variation} shows the variation of the EOS for different combinations of $f$ and $b$ and their impact on the dark energy density. Here, the values of $f$ and $b$ are chosen within a range that exhibits an underdamped oscillatory behavior, as this scenario has been found to be largely consistent with the majority of supernova compilations.
	In our current framework, we restrict ourselves to the case in which the equilibrium point is around $w_m=-1$, as this case has been discussed extensively in the literature on oscillating EOS models \cite{Rezaei:2024vtg,Escamilla:2024fzq} and has also been found to have higher Bayesian evidence compared to both the $\Lambda$CDM and CPL models for the Pantheon+ sample \cite{Hussain:2026srf}.
	
	The Hubble equation can then be written as
	\begin{equation}
		E^2 = \left((1-\Omega_{\rm m0}-\Omega_{\rm r0})\ f_{\rm de} + \Omega_{\rm m0}\ e^{-3 N} + \Omega_{\rm r0}\ e^{-4 N} \right),
	\end{equation}
	where the dark energy density function is given by
	\begin{equation}
		f_{\rm de} = \exp\left(-3 \int_{0}^{N} (1+w_{\rm de}) dN \right)\ .
	\end{equation}
	The corresponding deceleration parameter and effective equation of state of the system are defined as
	\begin{equation}
		q = -1-\frac{\dot H}{H^2}, \ w_{\rm eff} = -1 - \frac{2 \dot H}{3H^2}\ . 
	\end{equation}
	
	We perturb the homogeneous and isotropic background line element as
	\begin{equation}
		ds^2 = a(\eta)^2\bigg[-(1+ 2 \psi)^2 d\eta^2+ (1- 2 \varphi)\delta_{ij}dx^i dx^j \bigg]\ ,
	\end{equation}
	where $\psi(\eta, \vec{x})$ and $\varphi(\eta, \vec{x})$ are the scalar potentials and $\eta$ denotes the conformal time. Since the considered model is dynamical, the perturbed dark energy density and velocity evolution in the Newtonian gauge are given by \cite{Ma:1995ey}
	\begin{align}
		\delta'_{\rm de} &= -(1+w_{\rm de}) (\theta - 3 \varphi')-3 \mathcal{H}(c_a^2 -w_{\rm de}) \delta \ , \label{delta_evol} \\
		\theta'_{\rm de} & = - \mathcal{H}(1-3w_{\rm de})\theta - \frac{w'_{\rm de}\theta}{1+w_{\rm de}} + \frac{c_a^2 k^2 \delta}{1+w_{\rm de}}  + k^2 \psi. \label{theta_evol}
	\end{align}
	Here $()' \equiv d()/d\eta$, and $\mathcal{H} = a'/a$, while $\delta_{\rm de} \equiv (\rho_{\rm de} - \bar{\rho}_{\rm de})/\bar \rho_{\rm de}$ denotes the density contrast of the dark energy component, with an overbar representing the background quantity. The quantity $\theta_{\rm de} \equiv \nb_{i}v^{i}$ denotes the velocity divergence, and $k$ denotes the perturbation mode in Fourier space. The corresponding adiabatic sound speed $\ca$ is given by
	\begin{multline}
		\ca \equiv \frac{\delta P}{\delta \rde}  = \frac{d(w_{\rm de}\rde)}{d \rde}  = w_{\rm de} + \rde \frac{d w_{\rm de}}{d a} \frac{da }{d \rde}\ .
	\end{multline}
	As the current model crosses $w_{\rm de}=-1$, the above fluid perturbation
	equations become singular because of the factors of $(1+w_{\rm de})$ in the denominators. The \texttt{CLASS} code handles this crossing using the parametrized post-Friedmann (PPF) framework \cite{Hu:2008zd,Fang:2008sn}, in which the dark-energy density and momentum perturbations are replaced by a single dynamical variable $\Gamma$, from which $\delta_{\rm de}$ and $\theta_{\rm de}$ are subsequently reconstructed. The evolution equation for $\Gamma$ remains regular across the phantom divide, $w_{\rm de}=-1$, while preserving energy--momentum conservation. We employ the PPF prescription throughout our analysis.

	\section{Observational Samples and Methodology}
	\label{sec:obs_method}
	
	We constrain the model parameters using the publicly available Python framework for Markov Chain Monte Carlo (MCMC) sampling, \texttt{Cobaya} \cite{Torrado:2020dgo,Lewis:2002ah,Lewis:2013hha}, and a modified version of the Boltzmann solver \texttt{CLASS} \cite{Class_code} to implement the dark energy models. We use the Gelman--Rubin convergence diagnostic, requiring $R-1<0.01$ to assess the convergence of the MCMC chains. The posterior distributions are analyzed and plotted using \texttt{GetDist} \cite{Lewis:2019xzd}.
	
	We use the following datasets to constrain the model parameters:
	
	\begin{itemize}
		\item \textbf{Planck 2018:} We use the full Planck 2018 likelihoods, including the low-$\ell$ temperature and low-$\ell$ EE likelihoods for $2 \leq \ell \leq 29$, the high-$\ell$ \texttt{Plik} TT,TE,EE likelihood for $\ell>30$, together with the Planck CMB lensing likelihood \cite{Planck:2018nkj,Planck:2018vyg,Aghanim:2019ame,Aghanim:2018oex}. We refer to this combination of datasets as Planck.
		
		\item \textbf{DESI BAO:} We employ the Baryon Acoustic Oscillation (BAO) measurements from the Dark Energy Spectroscopic Instrument (DESI) Data Release II \cite{DESI:2025zgx}, which we refer to as DBAO.
		
		\item \textbf{Type Ia supernovae:} We utilize three independent Type Ia supernova compilations: Pantheon+ (PP), excluding the low-redshift samples with $z<0.01$ \cite{Brout:2022vxf}, resulting in an effective sample size of $1590$; the updated Dark Energy Survey (DES-Dovekie) sample \cite{DES:2025sig}; and the Union3 compilation \cite{Rubin:2023jdq}. Throughout this manuscript, we refer to these datasets as PP, DES, and Uni3, respectively.
		
	\end{itemize}
	
We obtain constraints on the model parameters using the combined likelihoods, considering each of the three independent Type Ia supernova compilations separately:
\begin{equation}
	\mathcal{L}_{\rm tot} = \mathcal{L}_{\rm Planck} \times \mathcal{L}_{\rm DBAO} \times \mathcal{L}_{\rm (PP, DES, Uni3)}\ .
\end{equation}
	
	\section{Results}
	\label{sec:result}
	
We obtain constraints on the model parameters by implementing the differential equation given in Eq.~\eqref{eq:osc} in the \texttt{CLASS} framework. We vary the model parameters and nuisance parameters with uniform priors over the ranges specified in Tab.~\ref{tab:prior}. The resulting parameter constraints at the $68\%$ confidence level are summarized in Tab.~\ref{tab:param_const}, while the marginalized posterior distributions of the model parameters are shown in the corner plot in Fig.~\ref{fig:dho_triangle}.
	\begin{table}[tbh]
		\centering
		\begin{tabular}{lr}
			\toprule
			Parameter & Range\\
			\toprule
			$\Omega_{\rm b}h^2$ & $[0,0.1]$\\
			$\ln(10^{10}A_s) $ & $[1.61, 3.91]$\\
			$n_s$ & $[0.8, 1.2]$\\
			$100\theta_s$ & $[0.5, 10]$\\
			$\Omega_\mathrm{c} h^2$ & $[0.001, 0.99]$\\
			$\tau_\mathrm{reio} $ & $[0.01,0.8]$\\
			$N_{\rm eff}$ & $[2.0,4.0]$\\
			$w_0$ &  $[-6.0,2.0]$ \\
			$w_a$ & $[-8.0,8.0]$\\
			$f$ & $[0.0,90.0]$\\
			$b$ & $[-0.48, 120]$\\
			$H_0$ & $[30,100]$\\		
			\bottomrule
			\bottomrule
		\end{tabular}
		\caption{The uniform prior range of the model parameters.}
		\label{tab:prior}
	\end{table}
	\begin{figure*}[tbh]
		\resizebox{\textwidth}{!}{\includegraphics{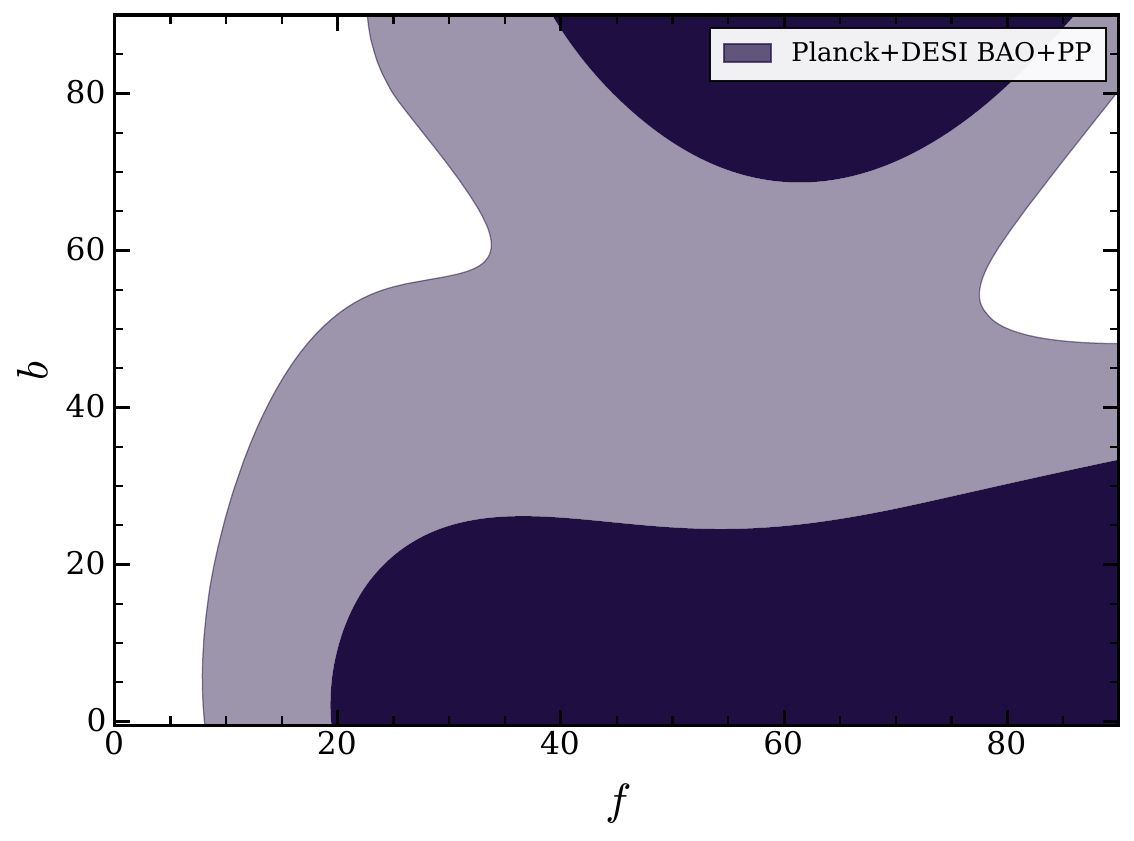}, \includegraphics{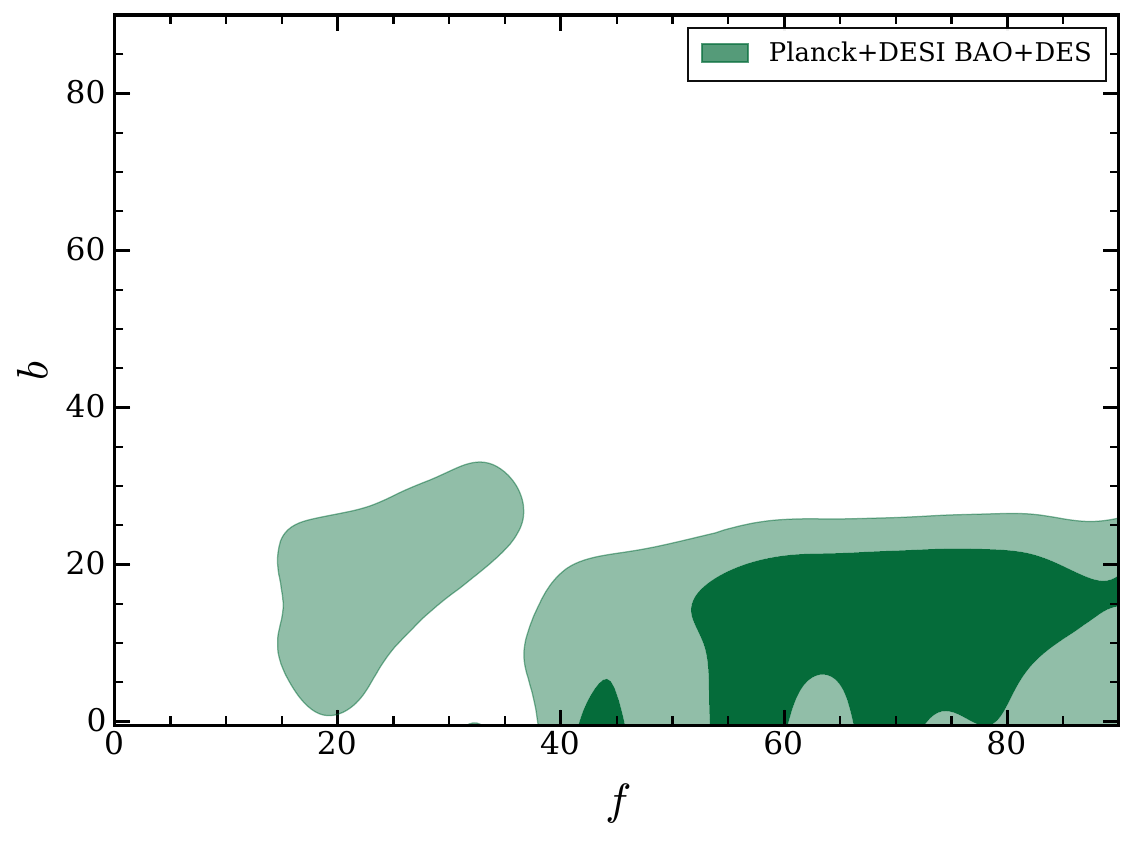}\includegraphics{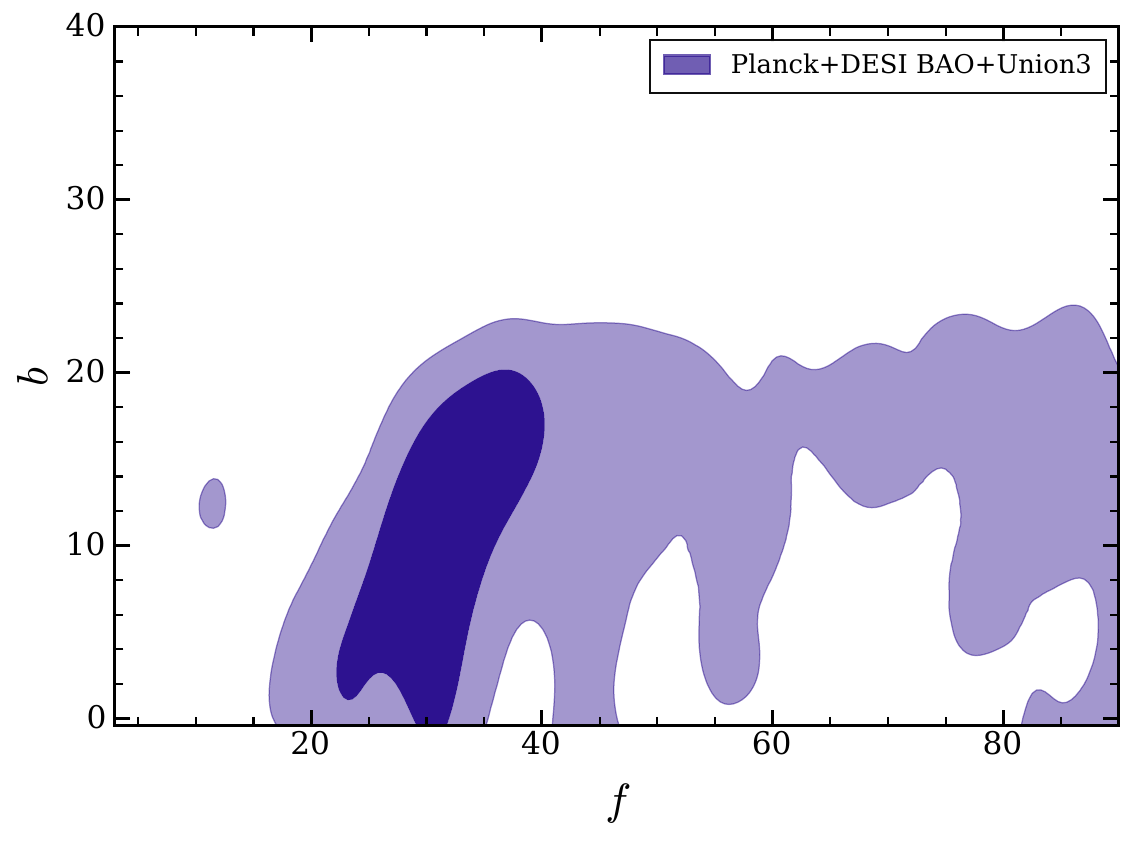} }
		\caption{The marginalized 2D posterior plot of the model parameters corresponding to $\cs =1$.}
		\label{fig:2dplot_fb}
	\end{figure*}
The posterior distributions in the $(f,b)$ parameter space for the three supernova compilations are shown in Fig.~\ref{fig:2dplot_fb}. We find that the posterior geometry in the damping parameter $b$ and frequency $f$ plane varies significantly among the different supernova compilations. For the Pantheon+ compilation, the posterior distribution of $b$ extends from lower to higher values, which differs noticeably from the distribution obtained using the CMB distance prior reported in Ref.~\cite{Hussain:2026srf}. We find that, for moderately high frequencies, $f>30$, both the built-in NDF15 and Runge--Kutta (RK) integrators in the \texttt{CLASS} code reduce their step sizes below the allowed minimum at lower redshifts, $z\lesssim 1$, corresponding to conformal times $\eta \gtrsim 2.5 \times 10^{3}$ Mpc, for $b>20$, and consequently fail to obtain a numerical solution. However, the integrators successfully evolve the system for $b>80$, which corresponds to the strongly overdamped regime. Importantly, the regions of parameter space in which the \texttt{CLASS} integration fails are not physically excluded. In particular, the underdamped high-frequency cases exhibit sharper transitions in the equation of state at low redshifts, causing the numerical evaluation of the PPF perturbations to break down. This parameter region is particularly relevant for the Pantheon+ compilation, for which the posterior geometry extends from low to high frequencies and includes the region with higher damping factors, $b>20$. In Fig.~\ref{fig:2dplot_fb}, this behavior is manifested as a large gray region around $20<b<75$, where the posterior probability density is extremely small, followed by a finite posterior density for $b>75$. In contrast, the posterior distributions obtained with the DES-Dovekie and Union3 compilations do not extend into the high-$b$ region. This numerical limitation therefore introduces a practical difficulty in fully exploring regions of the parameter space that may contain physically meaningful solutions, particularly for the Pantheon+ compilation and potentially for other supernova datasets.
	\begin{figure}[tbh]
		\centering
		\resizebox{\columnwidth}{!}{\includegraphics{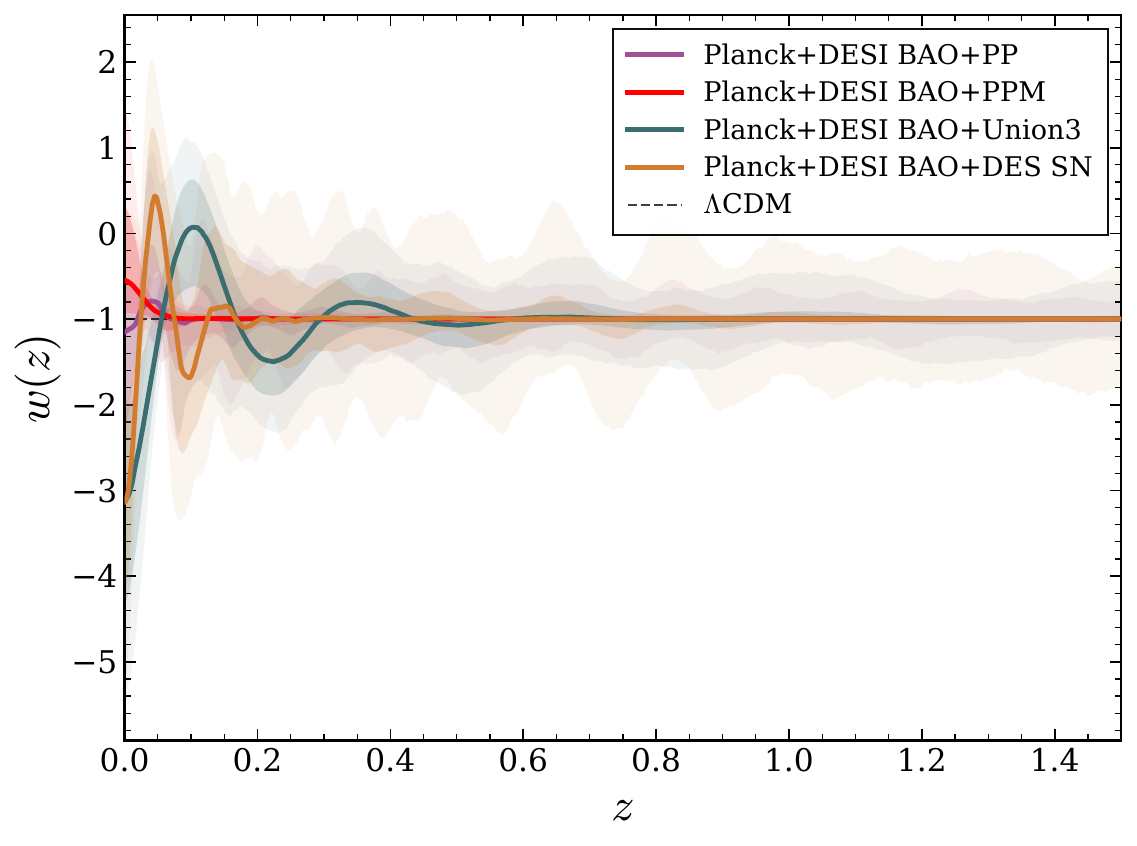}}
		\resizebox{\columnwidth}{!}{\includegraphics{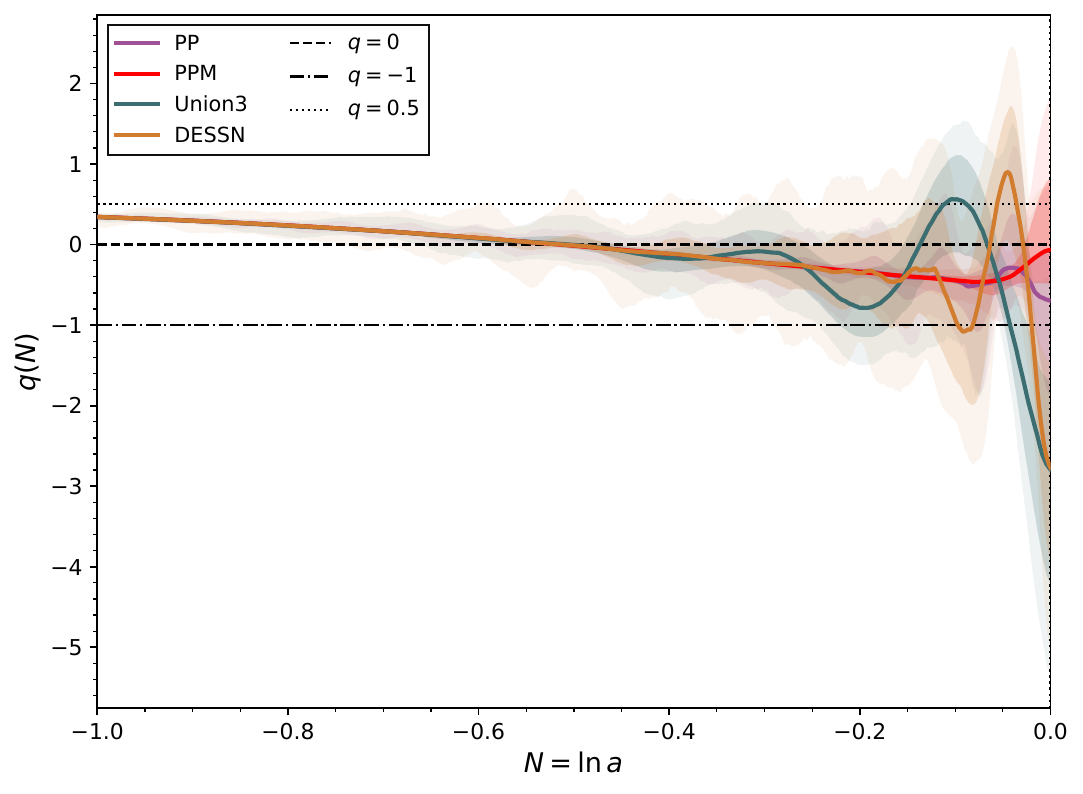}}
		\caption{Reconstructed equation of state for the DHO model corresponding to the different supernova compilations. The shaded regions indicate the $68\%$ and $95\%$ credible intervals obtained from the MCMC chains, while the solid lines denote the best-fit values reconstructed from the MCMC chains.}
		\label{fig:eos_dho}
	\end{figure}
	Hence, we address this problem by reducing the rest-frame dark-energy sound speed $(\cs)$ while retaining the PPF prescription. Specifically, we set $\cs \sim 10^{-5}$, whereas \texttt{CLASS} by default sets the sound speed to $\cs=1$, as the sound speed of a canonical scalar field is unity. { We emphasize that the choice $\cs=10^{-5}$ is adopted as a numerical prescription rather than as a physical assumption about the microphysics of dark energy. The DHO model is introduced phenomenologically at the level of the equation of state, and we do not assume a specific underlying field-theoretic realization that uniquely determines the rest-frame sound speed. The reduced value of $\cs$ is therefore used solely to alleviate the numerical stiffness encountered in the perturbation evolution in \texttt{CLASS}, allowing us to explore regions of the $(f,b)$ parameter space that are otherwise numerically inaccessible. 
		
		Although this prescription successfully enables \texttt{CLASS} to obtain solutions in the numerically pathological region of the $(f,b)$ parameter space, it may, in principle, introduce a scenario with clustering dark energy over a broad range of scales. Such clustering can significantly affect the CMB power spectrum at low multipoles, $\ell < 30$, through the Integrated Sachs--Wolfe (ISW) effect \cite{Das:2013sca}. Contrary to this expectation, however, we find no significant evidence for dark-energy clustering in the present analysis, as demonstrated below.
		
}


	\begin{figure*}[t]
		\resizebox{\textwidth}{!}{\includegraphics{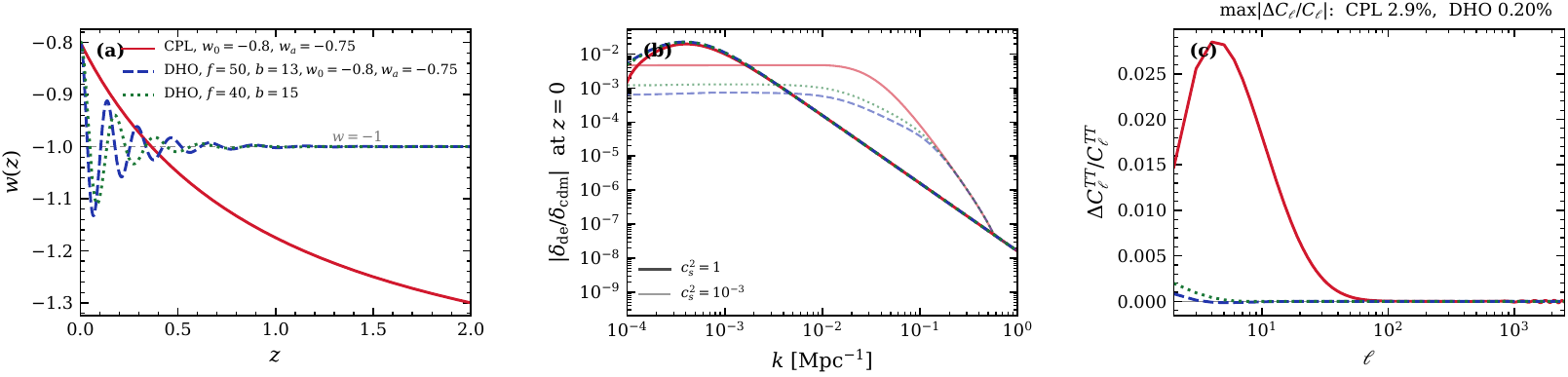}}
		\resizebox{\textwidth}{!}{\includegraphics{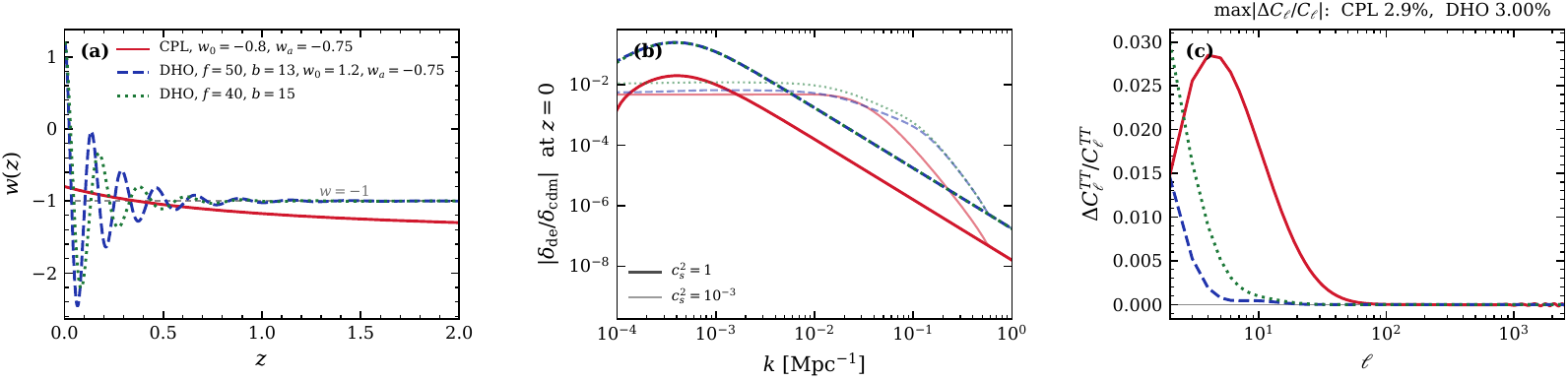}}
		\resizebox{\textwidth}{!}{\includegraphics{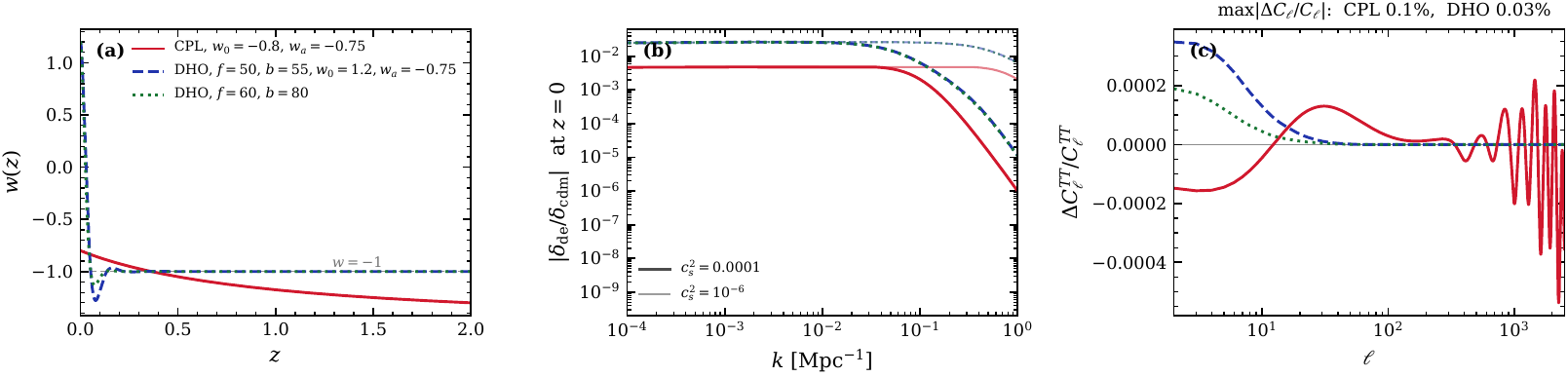}}
		\resizebox{\textwidth}{!}{\includegraphics{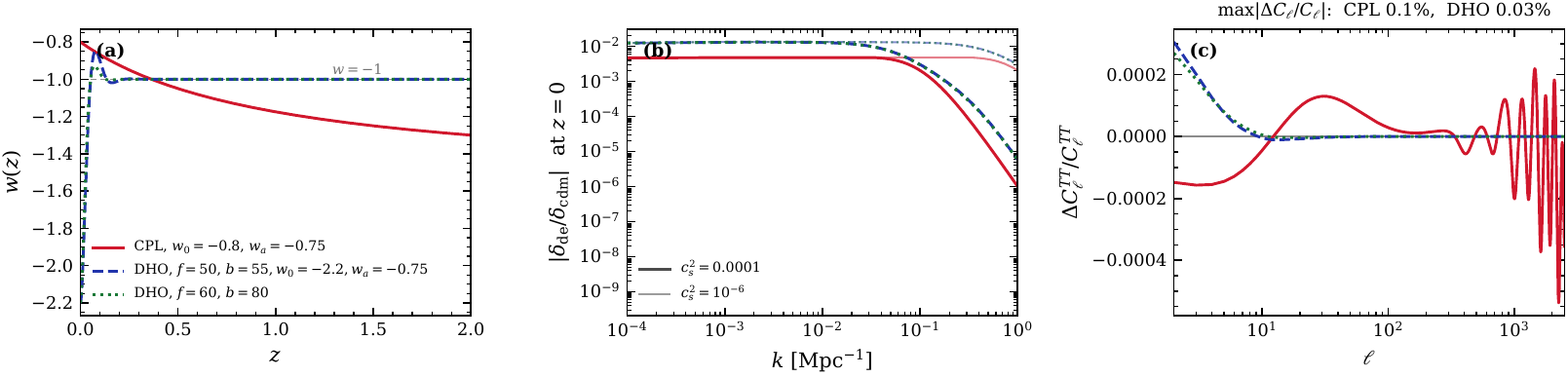}}
		\caption{Sensitivity test of the DHO and CPL models for different values of the sound speed, $\cs=(1,10^{-3},10^{-4},10^{-6})$, and for different values of the frequency, damping parameter, and equation of state. The sensitivity is assessed by evaluating the ratio $|\delta_{\rm de}/\delta_{\rm cdm}|$ at $z=0$ as a function of the wave number $k$, together with the lensed temperature power spectrum $C_{\ell}^{TT}$.}
		\label{fig:sound_speed_variation_impact}
	\end{figure*}

	\subsection{Impact of low sound speed}
	
Within the PPF framework, the perturbed dark energy density and momentum density are replaced by a single dynamical variable, $\Gamma$, whose equation of motion is constructed to preserve the conservation of energy and momentum \cite{Fang:2008sn}. Nevertheless, the evolution becomes numerically stiff at low redshifts for $\cs=1$, particularly when the equation of state undergoes rapid oscillations with a high frequency and a comparatively large damping factor. Therefore, we reduce the sound speed from $\cs=1$ to $\cs=10^{-5}$. A finite sound speed introduces a Jeans scale, $k_J = \mathcal{H}/c_{s}$, above which pressure support suppresses the growth of dark energy perturbations, while modes with $k<k_J$ can cluster. Consequently, reducing the sound speed allows dark energy to cluster over a broader range of scales, and for a slowly varying dark energy EOS, this change can have significant and potentially measurable consequences for the growth of structure and the evolution of late-time gravitational potentials \cite{dePutter:2010vy}. However, the same behavior does not arise for the present model. On small scales, $k^2 \gg \mathcal{H}^2$, the dark energy perturbations remain suppressed, and even in the limit of a negligible sound speed \cite{Batista:2021uhb}, the density contrast is given by
\begin{equation}
	\delta_{\rm de} = \frac{1+w_{\rm de}}{1-3w_{\rm de}}\delta_{\rm m}\ .
\end{equation}
Since, at higher redshifts and for larger $b$, the equation of state of the current model approaches the equilibrium point $w_{\rm de}=-1$, the dark energy perturbations are strongly suppressed regardless of the value of $\cs$. In contrast, for a slowly varying dark energy equation of state, such as CPL, which can deviate appreciably from $-1$ at higher redshifts, the dark energy perturbations can become comparable to the matter perturbations in the negligible-sound-speed limit. Such enhanced perturbations can affect the formation of structure and the CMB temperature power spectrum. To demonstrate this behavior, we plot the ratio $|\delta_{\rm de}/\delta_{\rm cdm}|$ at $z=0$ as a function of the wave number $k$, together with the lensed temperature power spectrum $C_{\ell}^{TT}$ for the CPL and DHO models for different sound speeds, $\cs=(1,10^{-3},10^{-4},10^{-8})$, and different frequencies in Fig.~\ref{fig:sound_speed_variation_impact}.

In the first and second rows, where the frequency is larger than the damping parameter $(b<25)$, the \texttt{CLASS} integrator encounters no stiffness for $\cs=1$. In these cases, the DE equation of state undergoes rapid oscillations at low redshifts and approaches $w_{\rm de}=-1$ at higher redshifts. In the first row, both models have similar values of the equation-of-state parameters and exhibit similar behavior in the density-contrast ratio shown in panel (b) at $z=0$ for $\cs=1$ and $10^{-3}$. However, for lower values of $\cs$, the change in the density-contrast ratio for the DHO model remains smaller on small scales than that for the CPL model. In panel (c), the change in the temperature power spectrum for the CPL model when $\cs$ is reduced from $1$ to $10^{-3}$ is approximately $\sim 3\%$, whereas for the DHO model it is negligibly small, at around $0.2\%$.

In the second row, we consider $w_0=1.2$ for the DHO model with the same frequency, while the CPL parameters are kept fixed. We find that the change in the temperature power spectrum for the DHO model is nearly $3\%$, indicating that the impact on the low-$\ell$ modes can be comparable to that of the CPL model. This demonstrates that even a substantially different choice of the present-day equation of state for the DHO model does not lead to a larger impact on the low-$\ell$ temperature power spectrum for the lower sound speed considered here.

In the third row, we consider a higher frequency and a large value of $b$, corresponding to a nearly critically damped solution. For this choice, the \texttt{CLASS} solver encounters numerical difficulties, and hence we reduce $\cs$. This case demonstrates that the resulting effect on the temperature power spectrum remains small, at approximately $\sim 0.03\%$, compared with $\sim 0.1\%$ for the CPL model. A similar behavior is observed in the fourth row, which corresponds to $w_0=-2.2$.

Consequently, we treat the reduction of the sound speed to $\cs = 10^{-5}$ not as a fundamental
physical modification of the dark energy sector, but as a necessary numerical regularization to bypass stiffness in the Boltzmann hierarchy during rapid phantom crossings. As
demonstrated in Fig. \ref{fig:planck_cl_power_spectra}, this prescription leaves the relevant cosmological observables,
including the ISW effect and matter power spectrum, invariant at the $\lesssim 10^{-3}$ level. We
caution, however, that this prescription is validated only for the specific parameter ranges
considered here.

 In the remainder of the paper, we therefore present parameter constraints for the DHO model evaluated with both $\cs=1$ and $\cs=10^{-5}$. Specifically, we report the results corresponding to both $\cs=10^{-5}$ and $\cs=1$ only for the Pantheon+ compilation. For the other datasets, we retain the results obtained with $\cs=1$. Henceforth, we denote the Pantheon+ results obtained with $\cs=10^{-5}$ by PPM (``Pantheon+, modified sound speed'') to distinguish them from the PP results obtained with $\cs=1$. We note, however, that the lower sound speed allows us to explore a larger range of the damping parameter $b$, as demonstrated in Fig.~\ref{fig:2dplot_fb}. Nevertheless, based on the sensitivity analysis presented above, we do not expect this additional parameter space to produce deviations larger than those reported here.
	
	\subsection{Parameter Constraints}
	
\begin{figure*}
		\resizebox{\textwidth}{!}{\includegraphics{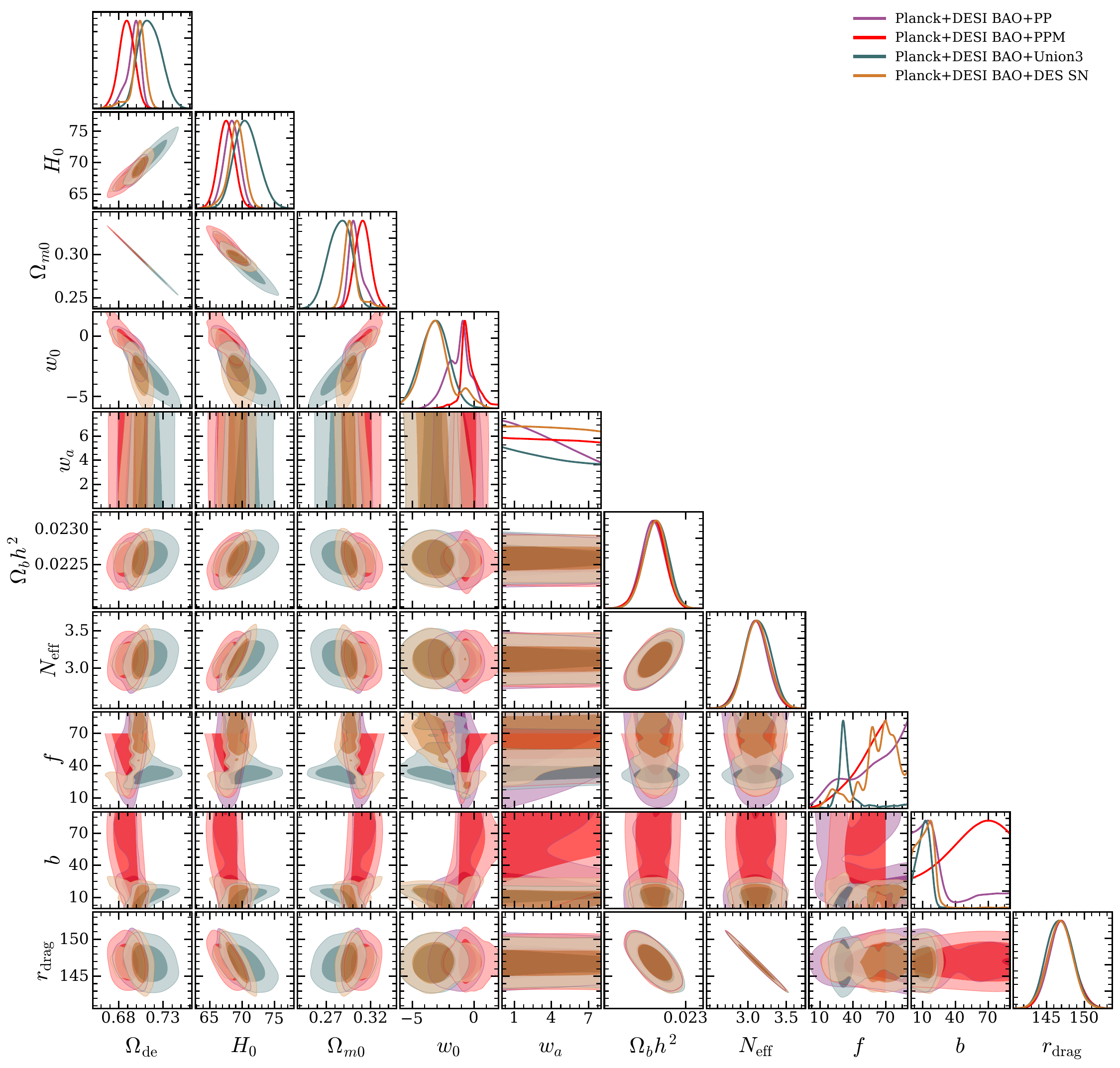}}
		\caption{Marginalized posterior distribution of the DHO model for the combinations of data sets.}
		\label{fig:dho_triangle}
	\end{figure*}

	The posterior distributions of all the model parameters are shown in Fig.~\ref{fig:dho_triangle}. For the DHO model, we vary the effective number of relativistic species, $N_{\rm eff}$, with a uniform prior, while for the CPL and $\Lambda$CDM models, it is fixed to $N_{\rm eff}=3.044$. The constraints on the cosmological and nuisance parameters for the DHO, CPL, and $\Lambda$CDM models are reported in Tab.~\ref{tab:param_const}.
	\begin{figure}
		\resizebox{\columnwidth}{!}{\includegraphics{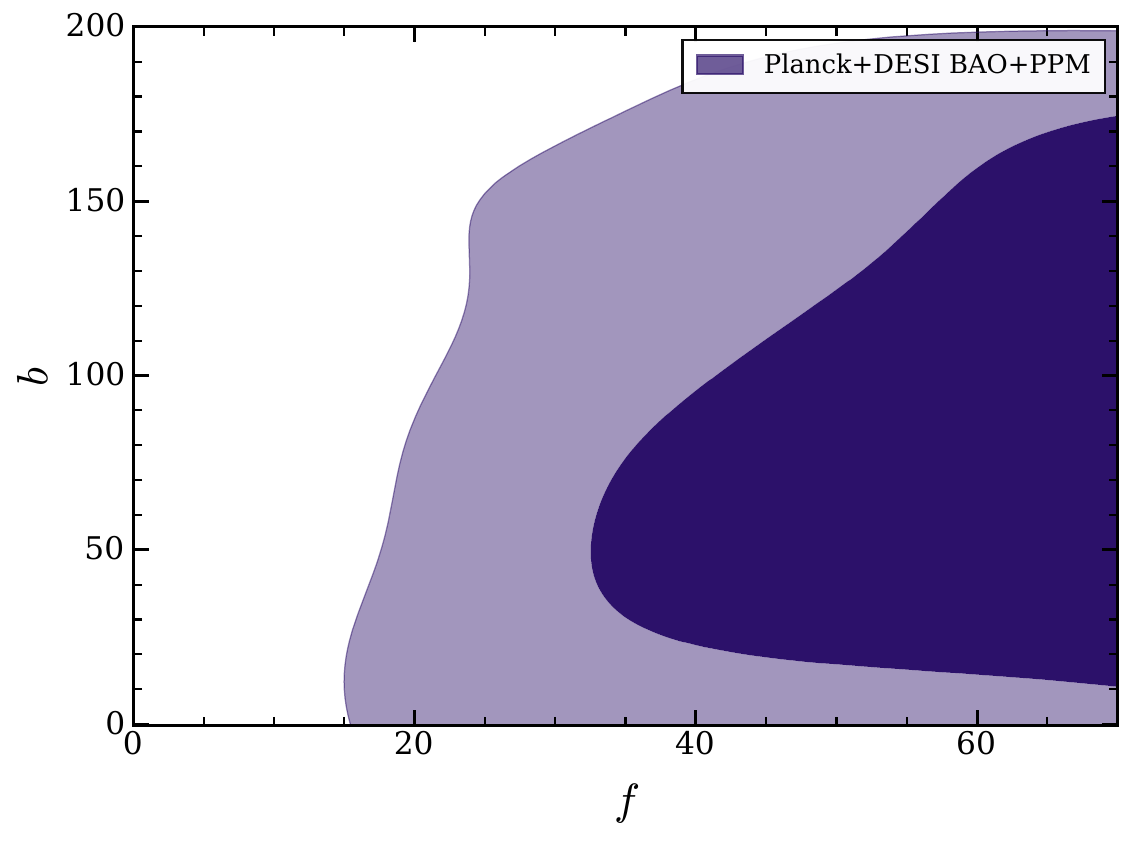}}
		\caption{Marginalized 2D posterior distribution of $f$ and $b$ for $\cs=10^{-5}$.}
		\label{fig:2d_fb_ppm}
	\end{figure}
	The two-dimensional posterior distribution in the $(f,b)$ parameter space for PPM is shown in Fig.~\ref{fig:2d_fb_ppm}. For $\cs=10^{-5}$, we find that the preferred values of $b$ increase with increasing $f$, resulting in an overdamped solution, in contrast to the posterior distribution shown in Fig.~\ref{fig:2dplot_fb}. Consequently, the corresponding equation of state exhibits no oscillatory behavior, as shown in Fig.~\ref{fig:eos_dho}. We also show the deceleration parameter, $q=-1-\frac{\dot H}{H^2}$. For PP, we find $q_0\sim-0.6$ and $w_0\sim-1$, whereas for PPM, $q_0\sim0$ and $w_0\sim-0.6$, indicating a significantly weaker late-time acceleration in the latter case. 
	The contrasting behavior between PP and PPM emerges only at very low redshifts, $z\lesssim0.1$. 
	\begin{figure*}[tbh]
		\resizebox{\textwidth}{!}{\includegraphics{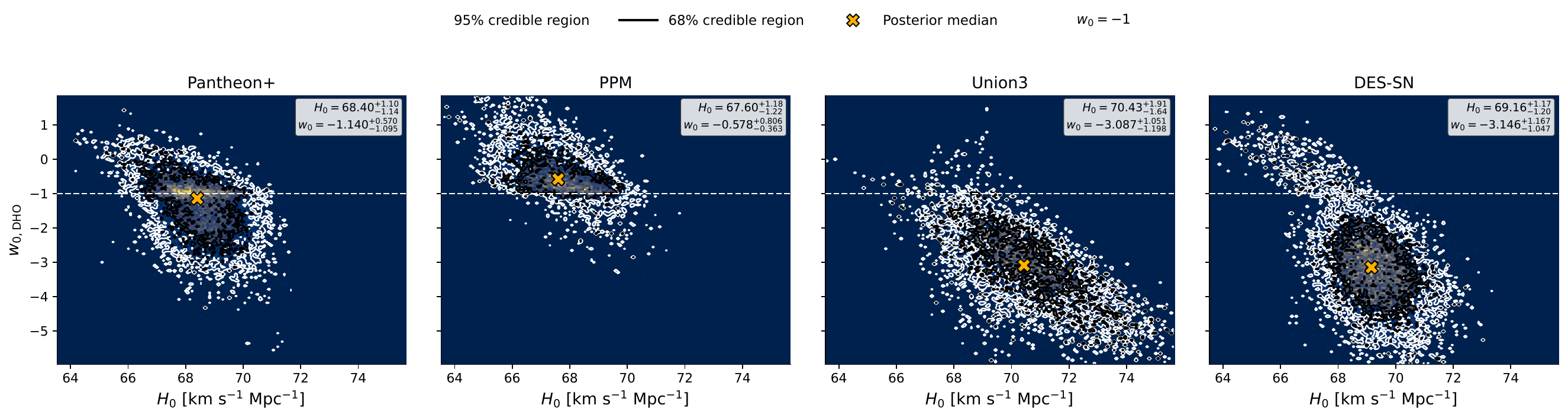		} }
		\resizebox{\textwidth}{!}{\includegraphics{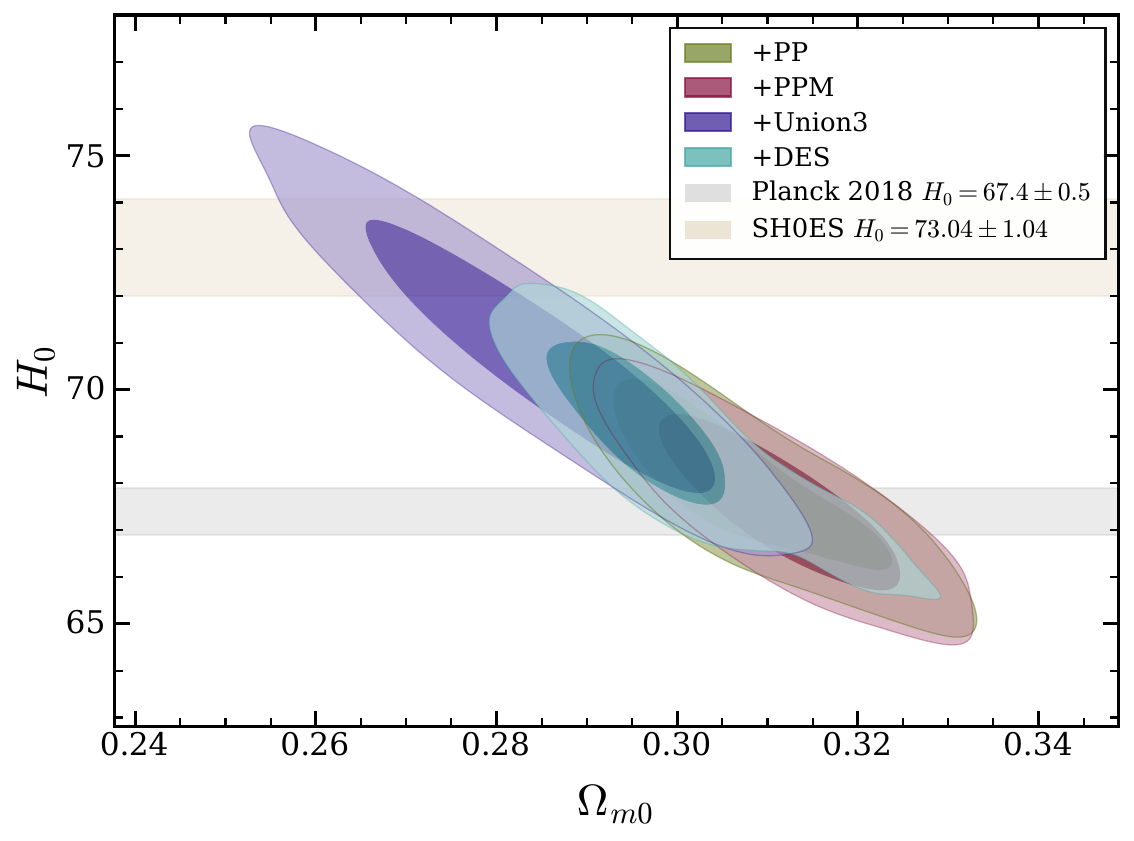}\includegraphics{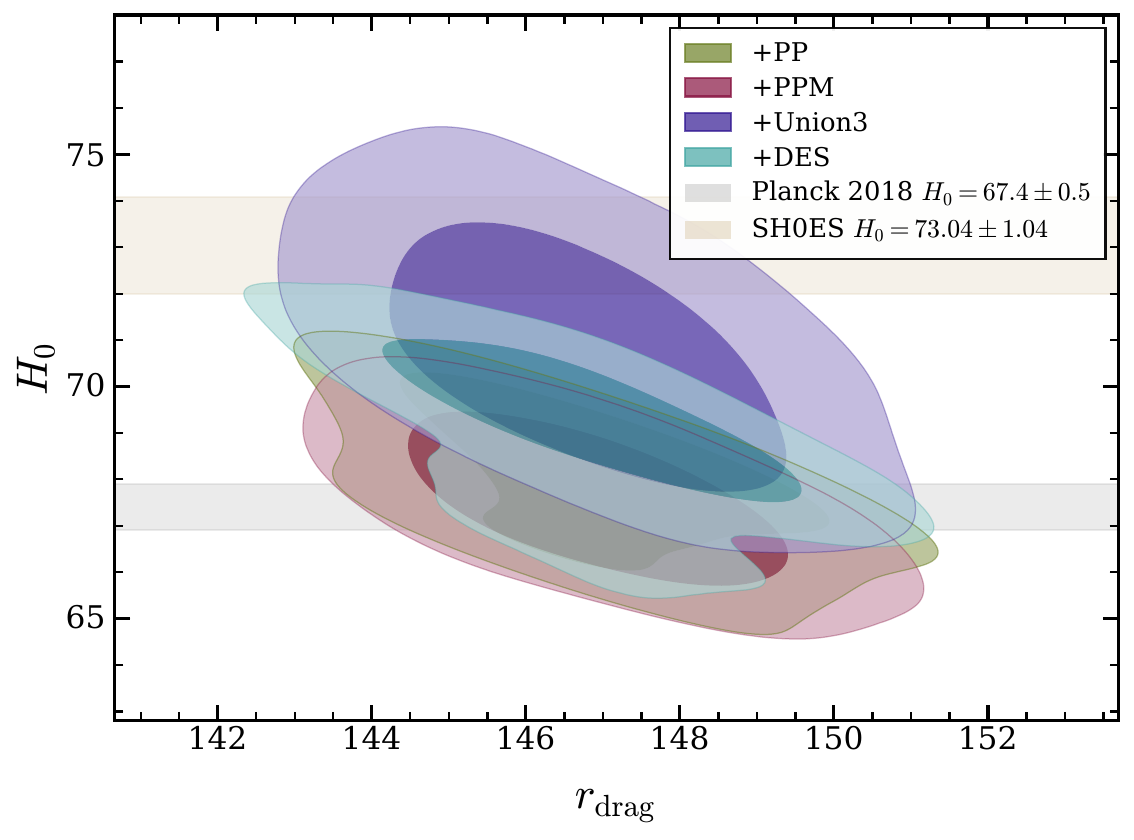} \includegraphics{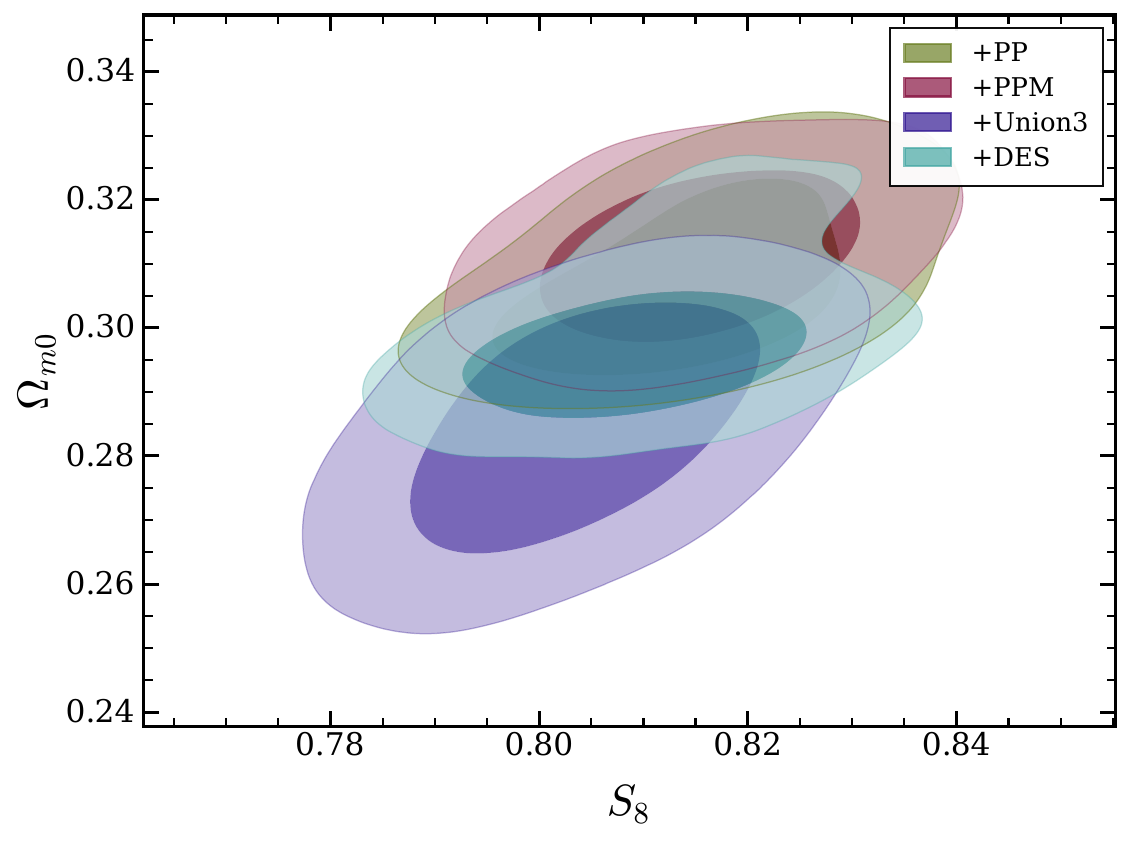}}
		\caption{Marginalized 2D posterior distributions for the cosmological parameter combinations $(H_0,w_0)$, $(\Omega_{\rm m0},H_0)$, $(r_{\rm drag},H_0)$, and $(S_8,\Omega_{\rm m0})$ for the DHO model. Planck+DESI BAO is common to all panels and serves as the baseline dataset combination. Here, PPM denotes the Pantheon+ analysis with $\cs=10^{-5}$, while PP denotes the Pantheon+ analysis with $\cs=1$.}
		\label{fig:2dplot_Hw0}
	\end{figure*}
	
Furthermore, we show the posterior distribution in the $(H_0,w_0)$ plane in Fig.~\ref{fig:2dplot_Hw0}. For PPM, the posterior density is concentrated around $w_0>-1$, with the best fit value $w_0=-0.578^{+0.806}_{-0.363}$ and $H_0\sim67.60$ \km, whereas PP yields $w_0=-1.140^{+0.57}_{-1.14}$ and $H_0\sim68.40$ \km. The constraints on $w_0$ differ by approximately $0.83\sigma$. For the PPM compilation, we obtain $\Omega_{\rm m0}\sim0.3113$, $N_{\rm eff}\sim3.11$, $S_8\sim0.8154$, and $r_d\sim146.9$ Mpc, as reported in Tab.~\ref{tab:param_const}. The posterior distributions of the other key cosmological parameters, except $w_0$, remain similar between PP and PPM, as shown in Fig.~\ref{fig:2dplot_Hw0}. Similarly, the constraints on the nuisance parameters reported in Tab.~\ref{tab:param_const} remain largely unchanged. However, PPM yields $\chi_{\rm min}^2=4190.95$, which is lower than the corresponding values for both PP and $\Lambda$CDM. We note that the difference in $\chi^2$ between PP and PPM is $\sim 1.2$, which is comparable to the scatter expected from the finite length of the chains. We therefore do not attach any significance to this difference.
	
The posterior distributions of $(f,b)$ for DES and Union3 are shown in Fig.~\ref{fig:2dplot_fb}. The constraints on $b$ lie predominantly in the region corresponding to underdamped oscillatory solutions, and the resulting evolution of the equation of state is illustrated in Fig.~\ref{fig:eos_dho}. The oscillatory behavior is confined to low redshifts, $z\lesssim0.2$ for DES and $z\lesssim0.6$ for Union3. Owing to the phase of the oscillation, the EOS reaches values as low as $w_0<-3.0$, indicating a strongly phantom regime. This behavior drives the Hubble parameter toward higher values than those obtained with both CPL and $\Lambda$CDM, yielding $H_0=69.08^{+1.23}_{-1.16}$ \km and $H_0=70.67\pm 1.88$ \km for DES and Union3, respectively, as shown in Fig.~\ref{fig:2dplot_Hw0}. Despite the higher values of $H_0$, the sound horizon remains largely unaffected, with $r_d\sim147$ Mpc. Consequently, $H_0$ exhibits a degeneracy with the matter density, with $\Omega_{\rm m0}\sim0.29$, such that a lower matter density accommodates a higher value of $H_0$ at essentially fixed sound horizon $r_d$. Therefore, the higher values of $H_0$ is the signature of purely geometric effect rather than the modification to the pre-recombination physics. Thus, these fits should not be interpreted as evidence for an alleviation of the Hubble tension with the SH0ES measurement. The $S_8$ values obtained from all the datasets remain consistent with those of $\Lambda$CDM, although a slightly higher value is obtained for CPL. Furthermore, the posterior distribution of $N_{\rm eff}$ is positively correlated with $H_0$ and negatively correlated with $r_d$, as shown in Fig.~\ref{fig:dho_triangle}. For all three supernova compilations, the model consistently yields $N_{\rm eff}\sim3.11$, with a slightly higher value for the Union3 compilation, corresponding to its higher preferred value of $H_0$. The obtained constraint on $N_{\rm eff}$ is consistent with the standard expectation, $N_{\rm eff}=3.044$, within $1\sigma$, showing no evidence for additional relativistic species in the context of this model.
	\begin{table*}
		\centering
		\renewcommand{\arraystretch}{1.2}
		
		\resizebox{\textwidth}{!}{%
			\begin{tabular}{lccccccc}
				\toprule
				\toprule
				Dataset & $\chi^2_{\min}$ & $\Delta\chi^2$ & $H_0$ & $\Omega_{\rm m}$
				& $w_0$ & $w_a$ & $f$ \hspace{4em}  $b$\\
				\midrule
				\multicolumn{8}{l}{\textit{\bf \boldmath Flat $\Lambda$CDM}} \\
				\;+PP      & $4193.9$ & $0.0$ & $68.33\pm0.29$ & $0.3020\pm0.0037$ & \textemdash & \textemdash & \textemdash \\
				\;+Union3  & $2816.7$ & $0.0$ & $68.30\pm0.29$ & $0.3024\pm0.0037$ & \textemdash & \textemdash & \textemdash \\
				\;+DES     & $4424.1$ & $0.0$ & $68.29\pm0.27$ & $0.3025\pm0.0035$ & \textemdash & \textemdash & \textemdash \\
				\midrule
				\multicolumn{8}{l}{\bf {CPL}} \\
				\;+PP      & $4186.9$ & $-7.0$  & $67.62\pm0.60$ & $0.3107\pm0.0057$ & $-0.842\pm0.054$ & $-0.58\pm0.21$ & \textemdash \\
				\;+Union3  & $2803.2$ & $-13.5$ & $66.01\pm0.82$ & $0.3269\pm0.0083$ & $-0.671\pm0.084$ & $-1.05\pm0.28$ & \textemdash \\
				\;+DES     & $4412.0$ & $-12.1$ & $67.46\pm0.55$ & $0.3126\pm0.0054$ & $-0.810\pm0.056$ & $-0.70\pm0.22$ & \textemdash \\
				\midrule
				\multicolumn{8}{l}{\textbf{DHO}} \\
				\;+PP      & $4192.1$ & $-1.8$ & $68.55\pm1.13$ & $0.3019\pm0.0064$
				& $-1.46_{-0.93}^{+0.89}$ & $3.6_{-2.6}^{+2.7}$
				& $60_{-28}^{+24}$ \hspace{0.6em} $16.8_{-12.2}^{+8.6}$ \\
				+PPM
				& $4190.95$
				& $-2.95$
				& $67.53_{-1.18}^{+1.22}$
				& $0.3113_{-0.0088}^{+0.0091}$
				& $-0.521_{-0.414}^{+0.891}$
				& $-0.264_{-5.396}^{+5.281}$
				& $54.47_{-19.02}^{+10.69}$ \hspace{0.6em}
				$80.15_{-41.03}^{+53.21}$ \\
				
				\;+Union3  & $2807.5$ & $-9.2$ & $70.67\pm1.88$ & $0.2850\pm0.0132$
				& $-3.16_{-1.16}^{+1.11}$ & $-1.0_{-5.1}^{+5.6}$
				& $36.6_{-8.3}^{+6.3}$ \hspace{0.6em} $11.3_{-6.4}^{+5.8}$ \\
				\;+DES     & $4414.3$ & $-9.8$ & $69.08_{-1.16}^{+1.23}$ & $0.2979_{-0.0075}^{+0.0059}$
				& $-3.01_{-1.21}^{+1.14}$ & $-0.1\pm5.5$
				& $62_{-17}^{+17}$ \hspace{0.6em} $13.5_{-9.3}^{+6.7}$ \\
				
				\bottomrule
				\bottomrule
		\end{tabular}}	
		
		\textbf{Primordial, clustering, and derived parameters}
		\resizebox{\textwidth}{!}{%
			\begin{tabular}{lcccccccccc}
				\toprule
				\toprule
				Dataset
				& $100\theta_s$
				& $\Omega_b h^2$
				& $\Omega_c h^2$
				& $N_{\rm eff}$
				& $r_{\rm d}$
				& $\ln(10^{10}A_s)$
				& $n_s$
				& $\tau_{\rm reio}$
				& $\sigma_8$
				& $S_8$ \\
				\midrule
				
				\multicolumn{11}{l}{\bf DHO} \\
				+PP
				& $1.04195_{-0.00046}^{+0.00047}$
				& $0.02255_{-0.00016}^{+0.00017}$
				& $0.1188_{-0.00027}^{+0.00028}$
				& $3.094_{-0.164}^{+0.163}$
				& $147.04_{-1.66}^{+1.70}$
				& $3.054_{-0.016}^{+0.016}$
				& $0.9718_{-0.0059}^{+0.0061}$
				& $0.060_{-0.007}^{+0.007}$
				& $0.8095_{-0.0113}^{+0.0111}$
				& $0.8119_{-0.0107}^{+0.0103}$ \\
				
				+PPM
				& $1.04192_{-0.00044}^{+0.00048}$
				& $0.02257_{-0.00017}^{+0.00015}$
				& $0.11881_{-0.00028}^{+0.00028}$
				& $3.110_{-0.162}^{+0.163}$
				& $146.88_{-1.65}^{+1.67}$
				& $3.0545_{-0.0160}^{+0.0158}$
				& $0.9725_{-0.0062}^{+0.0058}$
				& $0.0599_{-0.0070}^{+0.0075}$
				& $0.8005_{-0.0117}^{+0.0120}$
				& $0.8154_{-0.0099}^{+0.0106}$ \\
				
				+Union3
				& $1.04193_{-0.00047}^{+0.00048}$
				& $0.02259_{-0.00017}^{+0.00016}$
				& $0.1188_{-0.00028}^{+0.00029}$
				& $3.124_{-0.168}^{+0.172}$
				& $146.78_{-1.75}^{+1.70}$
				& $3.056_{-0.016}^{+0.016}$
				& $0.9736_{-0.0062}^{+0.0063}$
				& $0.061_{-0.007}^{+0.008}$
				& $0.8253_{-0.0156}^{+0.0161}$
				& $0.8039_{-0.0109}^{+0.0110}$ \\
				
				+DES
				& $1.04192_{-0.00047}^{+0.00048}$
				& $0.02258_{-0.00016}^{+0.00016}$
				& $0.1189_{-0.00027}^{+0.00028}$
				& $3.115_{-0.159}^{+0.159}$
				& $146.84_{-1.63}^{+1.62}$
				& $3.055_{-0.015}^{+0.015}$
				& $0.9727_{-0.0059}^{+0.0060}$
				& $0.060_{-0.007}^{+0.007}$
				& $0.8131_{-0.0114}^{+0.0116}$
				& $0.8100_{-0.0104}^{+0.0104}$ \\
				
				\midrule
				\multicolumn{11}{l}{\bf CPL} \\
				+PP
				& $1.04195_{-0.00027}^{+0.00028}$
				& $0.02243_{-0.00013}^{+0.00012}$
				& $0.11896_{-0.00090}^{+0.00089}$
				& \textemdash
				& $147.31_{-0.21}^{+0.22}$
				& $3.044_{-0.014}^{+0.014}$
				& $0.9669_{-0.0037}^{+0.0036}$
				& $0.056_{-0.007}^{+0.007}$
				& $0.8092_{-0.0091}^{+0.0092}$
				& $0.8234_{-0.0098}^{+0.0098}$ \\
				
				+Union3
				& $1.04191_{-0.00028}^{+0.00029}$
				& $0.02240_{-0.00014}^{+0.00013}$
				& $0.11930_{-0.00089}^{+0.00090}$
				& \textemdash
				& $147.25_{-0.22}^{+0.23}$
				& $3.042_{-0.014}^{+0.014}$
				& $0.9662_{-0.0038}^{+0.0038}$
				& $0.054_{-0.007}^{+0.007}$
				& $0.7983_{-0.0102}^{+0.0103}$
				& $0.8331_{-0.0100}^{+0.0102}$ \\
				
				+DES
				& $1.04193_{-0.00026}^{+0.00028}$
				& $0.02241_{-0.00013}^{+0.00014}$
				& $0.11916_{-0.00086}^{+0.00088}$
				& \textemdash
				& $147.27_{-0.22}^{+0.23}$
				& $3.043_{-0.015}^{+0.014}$
				& $0.9666_{-0.0036}^{+0.0036}$
				& $0.055_{-0.007}^{+0.008}$
				& $0.8093_{-0.0088}^{+0.0088}$
				& $0.8260_{-0.0095}^{+0.0099}$ \\
				
				\midrule
				\multicolumn{11}{l}{\boldmath{\bf Flat $\Lambda$CDM }} \\
				+PP
				& $1.04207_{-0.00026}^{+0.00027}$
				& $0.02251_{-0.00012}^{+0.00012}$
				& $0.11784_{-0.00065}^{+0.00063}$
				& \textemdash
				& $147.52_{-0.19}^{+0.19}$
				& $3.052_{-0.014}^{+0.014}$
				& $0.9698_{-0.0034}^{+0.0034}$
				& $0.060_{-0.007}^{+0.007}$
				& $0.8079_{-0.0061}^{+0.0060}$
				& $0.8106_{-0.0082}^{+0.0084}$ \\
				
				+Union3
				& $1.04207_{-0.00026}^{+0.00027}$
				& $0.02251_{-0.00012}^{+0.00012}$
				& $0.11791_{-0.00068}^{+0.00065}$
				& \textemdash
				& $147.50_{-0.19}^{+0.20}$
				& $3.051_{-0.014}^{+0.015}$
				& $0.9695_{-0.0033}^{+0.0032}$
				& $0.060_{-0.007}^{+0.007}$
				& $0.8076_{-0.0057}^{+0.0061}$
				& $0.8108_{-0.0081}^{+0.0082}$ \\
				
				+DES
				& $1.04208_{-0.00027}^{+0.00028}$
				& $0.02251_{-0.00012}^{+0.00012}$
				& $0.11794_{-0.00062}^{+0.00060}$
				& \textemdash
				& $147.49_{-0.19}^{+0.19}$
				& $3.051_{-0.015}^{+0.015}$
				& $0.9696_{-0.0032}^{+0.0031}$
				& $0.060_{-0.007}^{+0.007}$
				& $0.8078_{-0.0062}^{+0.0062}$
				& $0.8112_{-0.0085}^{+0.0082}$ \\
				
				\bottomrule
				\bottomrule
		\end{tabular}}	
		\caption{Marginalized cosmological constraints at the $68\%$ confidence level for flat $\Lambda$CDM, CPL, and DHO. All rows use Planck+DESI BAO as the baseline, combined with the independent supernova compilations Pantheon+ (PP), Union3, or DES-Dovekie (DES). The quantity $\Delta\chi^2 \equiv \chi^2_{\rm Model}- \chi^2_{\Lambda}$ is quoted relative to flat $\Lambda$CDM. PPM denotes the Planck+DESI BAO+Pantheon+ dataset combination with $\cs=10^{-5}$. We varied $N_{\rm eff}$ for DHO and held fixed to $3.044$ for CPL and $\Lambda$CDM.}
		\label{tab:param_const}
	\end{table*}

	\begin{figure}
		\resizebox{\columnwidth}{0.25\textheight}{\includegraphics{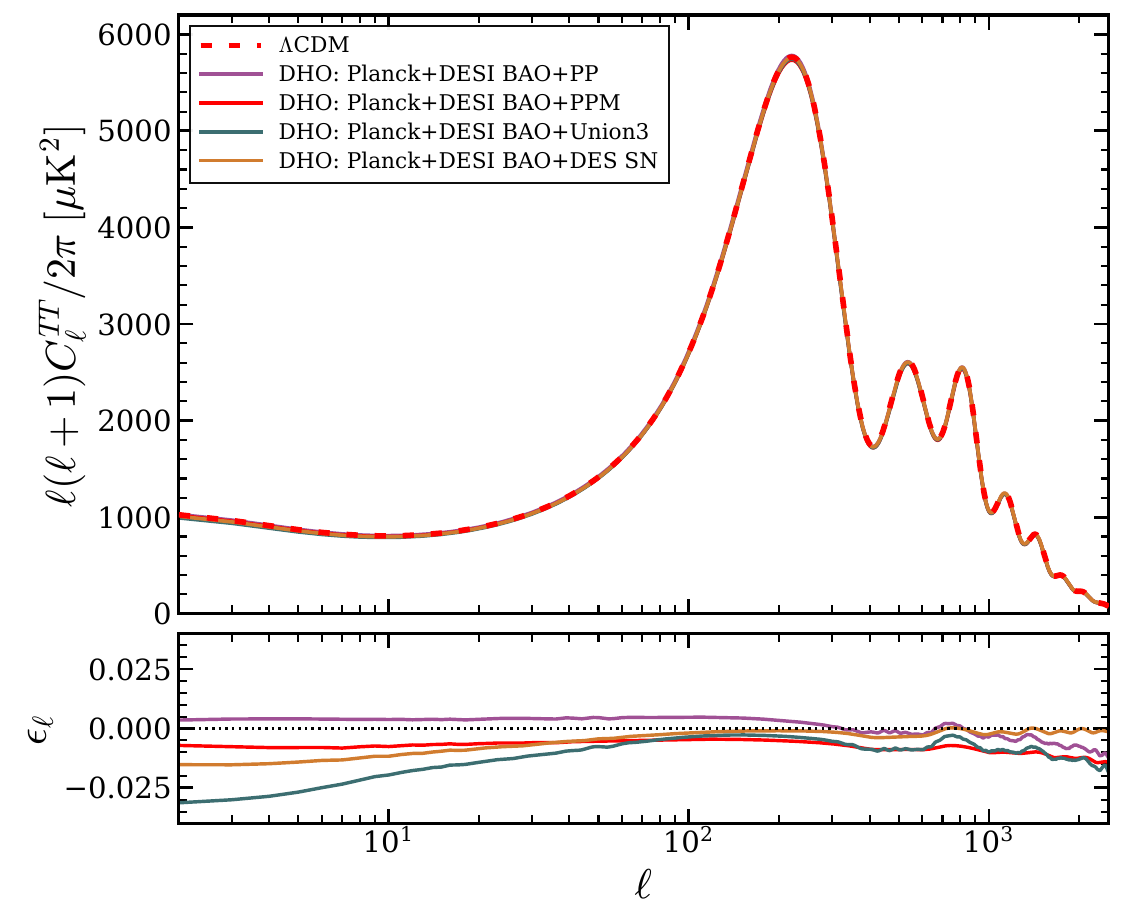}}
		\caption{CMB temperature power spectra, $C_{\ell}^{TT}$, for the DHO model compared with $\Lambda$CDM using the Planck 2018 likelihood for the considered combinations of datasets. The relative deviation, $\epsilon_\ell=\bigg(C_{\ell, \ \tiny{\rm DHO}}^{TT}/C_{\ell,\Lambda}^{TT}\bigg)-1$, is shown in the lower panel, indicating only a minute deviation from $\Lambda$CDM. At low $\ell$, the DHO model with PPM exhibits only a minute ISW contribution, indicating that the effect of dark-energy clustering becomes negligible for $\cs=10^{-5}$.}
		\label{fig:planck_cl_power_spectra}
	\end{figure}
	
	\begin{figure}
		\resizebox{\columnwidth}{0.3\textheight}{\includegraphics{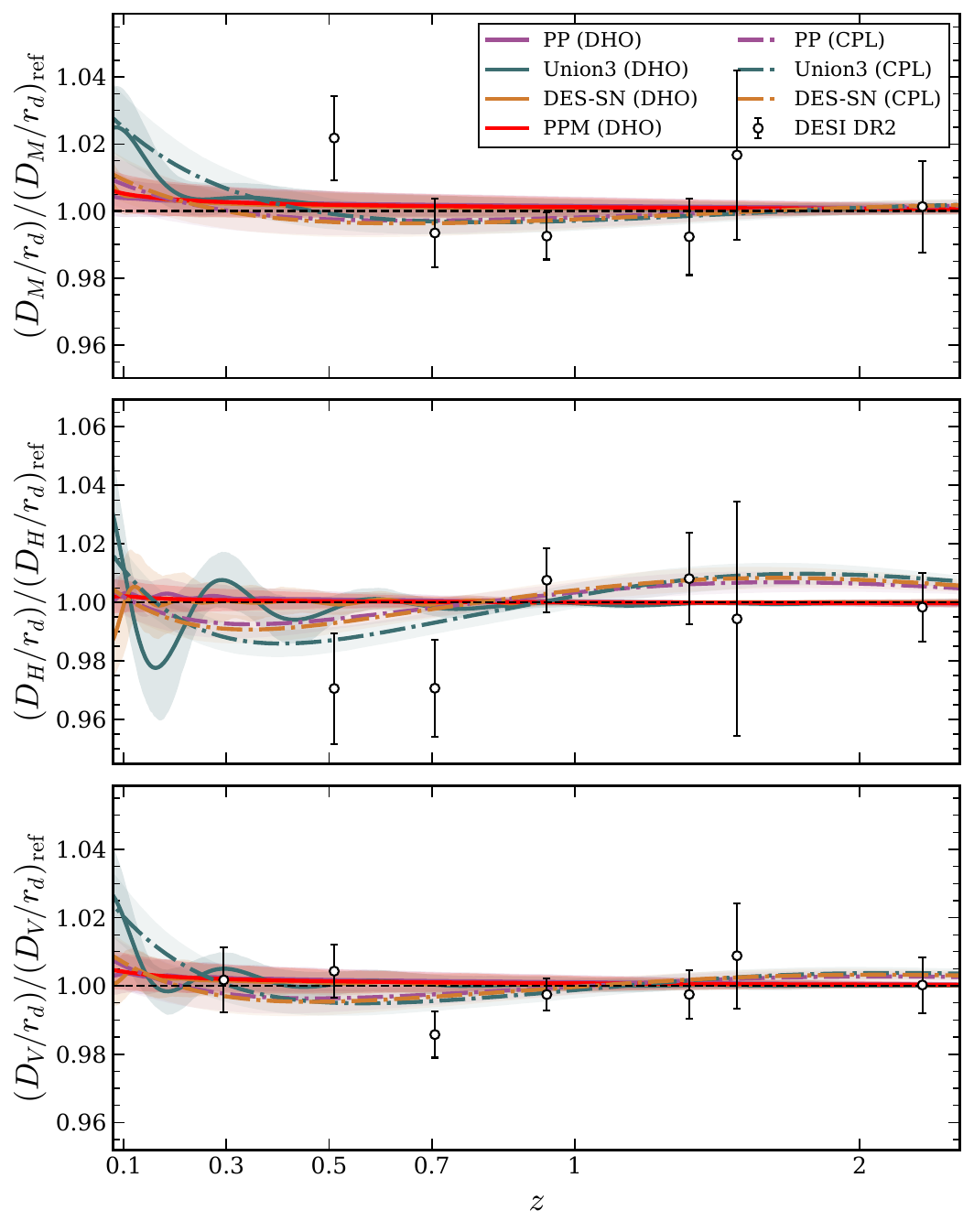}}
		\resizebox{\columnwidth}{0.65\textheight}{\includegraphics{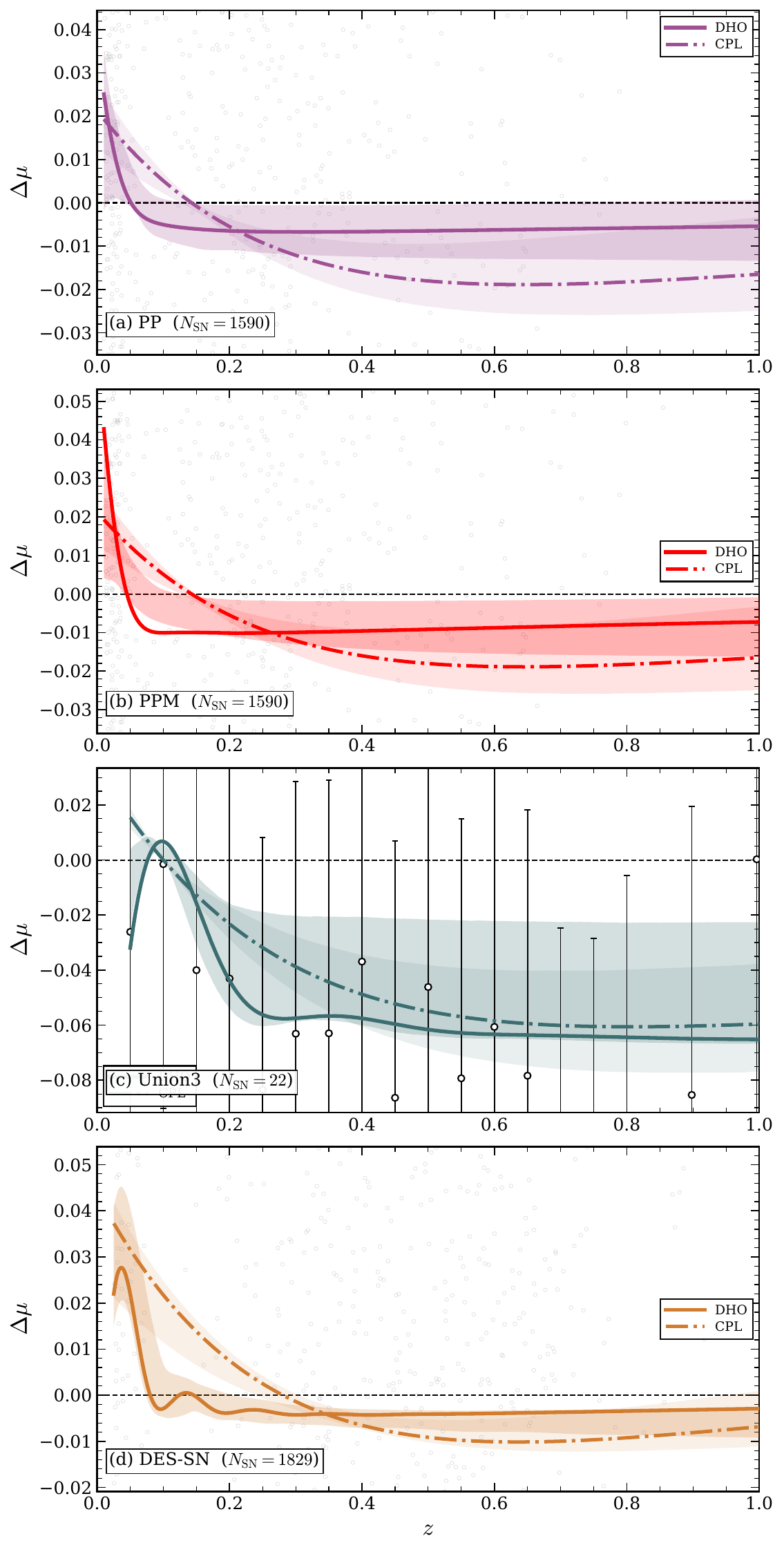}}
	\caption{Reconstructed geometric distances, $D_{M}/r_{d}$, $D_{H}/r_d$, and $D_{V}/r_d$, for DESI BAO for the DHO and CPL models relative to the fiducial model. The reconstructed distance modulus, $\Delta\mu \equiv \mu_{\tiny{\rm Model}}-\mu_{\Lambda}$, for each model is also shown relative to $\Lambda$CDM at the corresponding data points.}
		\label{fig:geometric_distance}
	\end{figure}
	
We plot the CMB temperature power spectra, $C_{\ell}^{TT}$, for the DHO model in Fig.~\ref{fig:planck_cl_power_spectra}. The model provides a fit similar to that of $\Lambda$CDM, with a deviation of less than $0.1\%$. For the PPM case, with $\cs=10^{-5}$, the deviation at low multipoles, $\ell<30$, also remains minute, indicating that the impact of dark-energy clustering on the CMB temperature power spectrum is negligible, as discussed in the previous section and shown in Fig.~\ref{fig:sound_speed_variation_impact}.

Additionally, we plot the DESI BAO geometric distances, $D_{M}/r_{d}$, $D_{H}/r_d$, and $D_{V}/r_d$, in Fig.~\ref{fig:geometric_distance}, where $D_M$ is the comoving angular diameter distance, $D_V$ is the spherically averaged distance, $D_H$ is the Hubble distance, and $r_d$ is the sound horizon at the drag epoch. In the $D_M/r_d$ panel, the DHO model produces an evolution similar to that of $\Lambda$CDM for PP, PPM, and DES. For Union3, however, DHO follows a behavior similar to CPL at $z<0.3$. At higher redshifts, $z>0.5$, the DHO prediction approaches that of $\Lambda$CDM, whereas the CPL prediction crosses the fiducial ratio of unity and approaches several of the data points.

In the $D_H/r_d$ panel, the oscillatory behavior of the DHO model appears only at $z<0.5$, where there are no corresponding data points for this quantity. This is an important limitation, as the redshift range in which the DHO model exhibits significant deviations is not probed by the BAO data. Consequently, the oscillatory signature is currently driven almost entirely by the low-redshift supernova data.
At higher redshifts, the oscillatory deviation is suppressed and the DHO prediction approaches the fiducial value. In comparison, the CPL prediction passes relatively close to the corresponding data points. In the $D_V/r_d$ panel, oscillatory features are visible for the Union3 compilation. The DHO prediction remains close to $\Lambda$CDM, while CPL exhibits a comparatively small deviation.

We also plot the relative distance-modulus difference, $\Delta\mu=\mu_{\rm Model}-\mu_{\Lambda{\rm CDM}}$, in Fig.~\ref{fig:geometric_distance}. The oscillatory behavior is most apparent for the DES and Union3 compilations, for which the DHO prediction deviates from $\Lambda$CDM by up to $1$--$3\%$ at $z<0.3$. For both PP and PPM, the DHO model follows a trend similar to that of CPL. The relatively large difference between the DHO and CPL predictions, particularly for DES and Union3 with PPM, is likely associated with the different distributions of supernovae at very low redshifts, $z\lesssim0.05$. Beyond this redshift, $z>0.05$, the predictions of the two dynamical models become very similar.

Finally, we evaluate the difference in the minimum chi-squared values, $\Delta\chi_{\rm min}^2$, for all models in Tab.~\ref{tab:param_const}. Relative to $\Lambda$CDM, the DHO model yields $\Delta\chi_{\rm min}^2=-9.8$, $-9.2$, and $-2.95$ for DES, Union3, and PPM, respectively, which is less favorable than the corresponding CPL fits, which yield $\Delta\chi_{\rm min}^2=-12.1$, $-13.5$, and $-7.0$. We further evaluate the Bayesian evidence using \texttt{MCEvidence} \cite{Heavens:2017afc}, with the results reported in Tab.~\ref{tab:bayes_evidence}. We define the logarithmic Bayes factor between models $i$ and $j$ as
\begin{equation}
	\Delta\ln B_{ij} = \ln Z_i - \ln Z_j,
\end{equation}
where $Z_i$ and $Z_j$ denote the Bayesian evidences of the respective models. A negative value of $\Delta\ln B_{ij}$ therefore indicates a preference for model $j$ over model $i$. Following the revised Jeffreys scale, $|\Delta\ln B|\lesssim1$, $1\lesssim|\Delta\ln B|<3$, $3\lesssim|\Delta\ln B|<5$, and $|\Delta\ln B|\gtrsim5$ correspond to inconclusive, moderate, strong, and very strong evidence, respectively. The Bayesian evidence consistently disfavors the DHO model relative to both $\Lambda$CDM and CPL. Compared with $\Lambda$CDM, we obtain $\Delta\ln B_{\rm DHO,\Lambda}=-5.76$, $-5.61$, $-2.57$, and $-2.03$ for PP, PPM, Union3, and DES, respectively, corresponding to very strong evidence for $\Lambda$CDM over DHO for PP and PPM, and moderate evidence for $\Lambda$CDM over DHO for Union3 and DES. Similarly, relative to CPL, we find $\Delta\ln B_{\rm DHO,CPL}=-3.40$, $-3.25$, $-4.49$, and $-1.41$ for PP, PPM, Union3, and DES, respectively, indicating strong evidence for CPL over DHO for PP, PPM, and Union3, and moderate evidence for CPL over DHO for DES. Thus, both the minimum $\chi^2$ and Bayesian evidence favor CPL over DHO for all the supernova compilations considered here.

Nevertheless, the distinct behavior of the DHO model across the three supernova compilations highlights the sensitivity of late-time cosmological inferences to the low-redshift supernova data. Although the model is not statistically favored over CPL, its ability to accommodate qualitatively different dynamical regimes provides a useful phenomenological framework for assessing whether the apparent preference for evolving dark energy is robust across independent supernova samples. In particular, the different damping and oscillatory behavior inferred from the DES-Dovekie, Union3, and Pantheon+ compilations suggests that improved and independent low-redshift supernova observations will be essential for determining whether these features reflect genuine late-time dynamics or residual differences among the supernova datasets. Future observations with larger and better-controlled samples may therefore provide a decisive test of oscillatory dark-energy phenomenology.

	\begin{table}
	\centering
	\begin{tabular}{lcc}
		\toprule
		Datasets & $\Delta \ln \rm B_{\rm DHO,\Lambda}$ & $\Delta \ln \rm B_{\rm DHO, CPL}$	\\
		\toprule
		{\bf	+PP }& $-5.76$ & $-3.40$ \\
		{\bf	+PPM} &  $-5.61$ & $-3.25$ \\
		{\bf	+Union3} & $-2.57$& $-4.49$\\
		{\bf	+DES} & $-2.03$ & $-1.41$ \\
		\bottomrule
		\bottomrule
	\end{tabular}
	\caption{The Bayesian evidence of the current model with respect to $\Lambda$CDM and CPL.}
	\label{tab:bayes_evidence}
\end{table}	
	
	\section{Conclusion}
	\label{sec:conclusion}
	
	In this paper, we have confronted a generalized oscillatory equation of state for dark energy with a combination of Planck, DESI BAO, Pantheon+, Union3, and DES-Dovekie supernova data. We treat Planck+DESI BAO as the baseline dataset and combine it independently with each of the three supernova compilations. 
	
	We implemented the DHO model in \texttt{CLASS} and found that, at the background level, its evolution is numerically robust over a well-defined region of the $(f,b)$ parameter space. At the level of linear perturbations, however, the evolution becomes numerically stiff in a regions of the parameter space where a model exhibits rapid oscillations. We overcome this difficulty by setting $\cs = 10^{-5}$, without introducing any significant dark-energy clustering on small scales. This is because, at higher redshifts, the equation of state approaches the cosmological-constant value, $w_{\rm de}=-1$, causing the dark-energy density perturbations to remain suppressed. We find that this prescription is necessary for the Pantheon+ compilation and therefore restrict its use to this dataset. For the remaining datasets, we adopt the default value $\cs = 1$.
	
	The oscillatory behavior of the model depends strongly on the damping parameter and, consequently, on the supernova compilation. For the Union3 and DES-Dovekie datasets, the preferred parameter regions correspond predominantly to underdamped solutions, resulting in pronounced oscillatory behavior at low redshifts. In contrast, the Pantheon+ compilation prefers a region with a larger damping parameter, corresponding to an overdamped solution and therefore suppressing the oscillatory behavior. This distinct response to different supernova compilations highlights the sensitivity of the DHO model to the low-redshift supernova data and provides a useful framework for investigating possible differences among the supernova samples.
	
	The present-day equation of state reaches strongly phantom values, $w_0<-3.0$, for the DES-Dovekie and Union3 compilations, whereas the Pantheon+ analysis favors a substantially less negative value, with $w_0>-0.6$. The different oscillatory behaviors are reflected in the inferred values of the Hubble constant, yielding $H_0=69.08^{+1.23}_{-1.16}$ \km and $H_0=70.67\pm 1.88$ \km for DES-Dovekie and Union3, respectively, while Pantheon+ (PPM) gives $H_0=67.53 ^{+1.22}_{-1.18}$ \km. The Pantheon+ result therefore remains in tension with the SH0ES measurement $(H_0 = 73.04 \pm 1.04)$ \km at approximately the $3.5\sigma$ level. Although the DES-Dovekie and Union3 analyses favor higher values of $H_0$, the sound horizon remains close to $r_d\sim147$ Mpc. The higher $H_0$ values are instead associated with a degeneracy with the matter density, $\Omega_{\rm m0}$, and therefore do not provide a compelling alleviation of the Hubble tension.
	
Overall, the DHO model yields $\chi^2_{\rm min}$ values intermediate between those of CPL and $\Lambda$CDM for the Union3 and DES-Dovekie compilations, while the logarithmic Bayes factors indicate strong evidence in favor of $\Lambda$CDM and moderate to strong evidence in favor of CPL over the DHO model. Thus, although the DHO model is not statistically preferred over CPL, its distinct response to different supernova datasets and its ability to produce qualitatively different late-time behaviors make it a useful phenomenological framework for testing the sensitivity of cosmological inferences to low-redshift observations. These results motivate further investigation of late-time oscillatory dark-energy models with more general functional forms and improved low-redshift supernova datasets.

	\begin{acknowledgements}
		S.H. acknowledges the support of the National Natural Science Foundation of China under Grants No. 12275238, No. 12542053, and No. W2433018, and the National Key Research and Development Program of China under Grant No. 2020YFC2201503. M.Z is supported by the National Key R\&D Intergovernmental Cooperation Program of China (2023YFE0102300,2024YFA1611500).
	\end{acknowledgements}

\bibliographystyle{apsrev4-2}

\bibliography{reference}
\end{document}